\documentclass[a4paper,fleqn,usenatbib]{mnras}

\usepackage[T1]{fontenc}
\usepackage{ae,aecompl}

\usepackage{graphicx}	
\usepackage{amsmath}	
\usepackage{amssymb}	

\title[The role of environment on the SFH of disc galaxies]
{The role of environment on the star formation history of disc galaxies}

\author[X. Y. Kang et al. ]
{Xiaoyu~Kang$^{1,2,3}$\thanks{E-mail: kxyysl@ynao.ac.cn}, Fenghui~Zhang$^{1,2,3}$ and Ruixiang  Chang$^{4}$\\
$^1$Yunnan Observatories, Chinese Academy of Sciences, 396 Yangfangwang, Guandu District, Kunming, 650216, P.R. China\\
$^2$Key Laboratory for the Structure and Evolution of Celestial Objects, Chinese Academy of Sciences, 396 Yangfangwang,\\
Guandu District, Kunming, 650216, P. R. China\\
$^3$Center for Astronomical Mega-Science, Chinese Academy of Sciences, 20A Datun Road, Chaoyang District, Beijing,\\ 100012, P.\,R.\,China\\
$^4$Key Laboratory for Research in Galaxies and Cosmology, Shanghai Astronomical Observatory, Chinese Academy of \\
Sciences, 80 Nandan Road, Shanghai, 200030, China
   }

\date{Accepted XXX. Received YYY; in original form ZZZ}

\pubyear{2016}

\begin{document}
\label{firstpage}
\pagerange{\pageref{firstpage}--\pageref{lastpage}}
\maketitle

\begin{abstract}

NGC\,2403, NGC\,300 and M33 are three nearby pure-disc galaxies with similar
stellar mass in different environments, they are benchmarks for understanding
late-type spiral galaxies in different environments. The chemical evolution and
growth of their discs are investigated by using the simple chemical evolution model,
in which their discs are assumed to originate and grow through the accretion of
the primordial gas, and the gas outflow process is also taken into account.
Through comparative study of the best-fitting model predicted star formation
histories for them, we hope to derive a picture of the local environment on
the evolution and star formation histories of galaxies and whether or not
the isolated galaxies follow similar evolution history. Our results show
that these three galaxies
accumulated more than 50 percent of their stellar mass at $z\,<\,1$.
It can be also found that the metallicity gradients in isolated spiral
galaxies NGC\,2403 and NGC\,300 are similar and obviously steeper
than that in M33, when the metallicity gradients are expressed in
${\rm dex}\,R_{\rm 25}^{-1}$. The similar metallicity gradients
in NGC\,2403 and NGC\,300 indicate that they may experience
similar chemical evolutionary histories. The principal epoch
of star formation on the discs of NGC\,2403 and NGC\,300 is earlier
than that on the disc of M33, and the mean age of stellar populations
along the discs of both NGC\,2403 and NGC\,300 is older than that of M33.
Our results indicates that the evolution and star formation history of a galaxy
indeed depends on its local environment, at least for galaxies with stellar
mass of $10^{9.2}\,\rm M_{\odot}\sim10^{9.7}\,\rm M_{\odot}$.
\end{abstract}




\section{Introduction}

Understanding the effect of environment on galaxy formation and evolution
is a central topic in extragalactic astronomy \citep{Gupta2016}.
An influence of environment on the evolution of late-type galaxies has been
studied by many groups, but no unanimous conclusion has been reached.
Some studies thought that galaxy environment has an important impact on the physical
properties of galaxies \citep{Spitzer1951, Davis1976}. For instance, compared
to field galaxies, the clustered spiral galaxies are appeared to be redder
\citep{Dressler1980, Kennicutt1983}, and the evolution timescale from
blue star-forming to red passive galaxies is shorter in dense environments
\citep{Wilman2005, Cooper2006, Poggianti2006, Iovino2010}. Spiral galaxies
in the cluster also have systematically higher metallicity than
comparable field galaxies \citep{Shields1991, Cooper2008}. However,
many recent studies suggest the effect of the environment is much weaker than
previously claimed. The evolution of late-type galaxies is driven mainly by their
intrinsic properties and is largely independent of their environment
over a large rang of local galaxy densities
\citep[][and references therein]{Mouhcine2007, Ellison2009, hughes13, Goddard2016,Pilyugin2016}.
Therefore, there exists some disagreement about whether or not the environment
plays a significant role in shaping the evolution of galaxies.

On the other hand, isolated spirals are characterized by similar abundance gradients,
regardless of the morphology, incidence of bars, absolute magnitude and mass,
when normalized to either the exponential disc scale-length $r_{\rm d}$ or the
isophotal radius ($R_{\rm 25}$)
\citep{zaritsky94, Garnett1997, Henry1999, PB2000, MM11, pilyugin14_1, Sanchez2014, ho15}.
Since the gas-phase oxygen abundance provides a fossil record of the formation
and evolution history of galaxies, the existence of a universal metallicity gradient
implies that all isolated disc galaxies may follow very similar chemical evolution
when growing their discs.

Individualized and comparative studies of the local spiral galaxies may help us understand
the environment effect on the evolutionary history of galaxies and whether
or not the isolated disc galaxies experience similar chemical evolution.
However, most discs are difficult to study in detail due to the
complexities of their bulges \citep{B&F95}. Fortunately, there are
three undisturbed and bulgeless (pure-disc) spiral galaxies (NGC\,2403,
NGC\,300 and M33) with different environments
\citep[see Figure 1 of][]{williams13a} in the nearby universe
($D<4\,\rm Mpc$). Due to the proximity, large angular sizes and low
inclinations for them, they are excellent targets for detailed observations
of their cold gas, star formation rate (SFR), metallicity and stellar
populations, thus providing abundant observational constraints for the
model of galaxy formation and evolution. In addition, these three galaxies
are similar in stellar mass (i.e., luminosity) and morphology, and
their basic observational properties are summarized in Table \ref{Tab:obs1}.

Among the three members, NGC\,2403 is the most isolated one, and it is
an outlying member of the M81 group. Its H{\sc i} rotation curve is
extremely regular \citep{Blok2008}, showing no warps \citep{williams13a}.
NGC\,300 is a relatively isolated member on the Sculptor filaments with
a nearby low-mass companion (NGC\,55), and its outer H{\sc i} disc shows
a severe warp \citep{puche90}, but there are no strong commonalities
between NGC\,300 and NGC\,55 \citep{Hillis2016}. Thus, NGC\,300 is undisturbed
and nearly the same as the isolated disc galaxy NGC\,2403 \citep{Hillis2016}.
Compared to the above mentioned two objects, M33 is the least isolated
one with a warped H{\sc i} disc between its disc and
M31 \citep{karachentsev04, putman09}, indicating they interacted with
each other in the past \citep{braun04, Bernard12}, and the H{\sc i} gas
streams exist along the axis between them, which are potential fuel
for future star formation in M33 and M31 \citep{Wolfe2013}.
Above all, the stellar surface
brightness profiles of both NGC\,2403 and NGC\,300 follow a pure
exponential form, while that of M33 has a break at about $8\,\rm kpc$
\citep{bland05, Ferguson07, barker07, barker11, barker2012, Hillis2016}.
All the aforementioned information indicates that NGC\,2403, NGC\,300
and M33 are ideal laboratories to explore the local environment on
the evolution of galaxies and whether or not the isolated spiral galaxies
follow the similar chemical evolutionary histories.
Thus, it is necessary to make a comparative study of the evolution and
star formation history (SFH) of these three disc galaxies.

\begin{table*}
\caption{Basic properties of NGC\,2403, NGC\,300 and M33.}
\label{Tab:obs1}
\begin{center}
\begin{tabular}{llll}
\hline
\hline
Property         &     NGC\,2403                 &   NGC\,300                     &   M33\\
\hline
Morphology       &     SAB(s)cd$^{\rm a, b}$     &  SA(s)d$^{\rm a, b}$           & SA(s)cd$^{\rm a, b}$  \\
Distance(Mpc)             &     3.2$^{\rm c,d}$          &  2.0$^{\rm c}$                 & 0.8$^{\rm e}$  \\
$M_{\rm B}(\rm mag)$      &     $-18.6^{\rm f}$          &  $-17.66^{\rm g}$              & $-18.4^{\rm g}$  \\
$M_{\rm K}(\rm mag)$      &     $-21.3^{\rm h} $          &  $-20.1^{\rm h}$            & $-20.4^{\rm h}$  \\
Scale-length\,(kpc)       &     1.6$^{\rm i}$             &  1.29$^{\rm j}$             &  1.4$^{\rm j}$   \\
Rotation velocity($\rm km\,s^{-1}$)     &   136$^{\rm k}$     &  91$^{\rm k}$           &  110$^{\rm k}$  \\
\hline
\end{tabular}\\
\end{center}
Refs: (a) NED; (b) \citet{1991S&T....82Q.621D}; (c) \citet{Dalcanton09};
(d) \citet{Freedman2001}; (e) \citet{Williams09}; (f) \citet{Lee2011};
(g) \citet{gogarten10}; (h) \citet{jarrett03}; (i) \citet{leroy08};
(j) \citet{MM07}; (k) \citet{garnett02}.
\end{table*}

The simple chemical evolution model is a fruitful tool to explore the
formation and evolution of disc galaxies \citep{Tinsley80}, and it has
been successfully applied to investigate the evolution and SFHs of nearby
disc galaxies
\citep[for instance,][]{chang99, chang12,chiappini01, PB2000, yin09, kang12, kang16}.
The goal of this work is to provide a picture of the local environment
on the evolution of galaxies and whether or not the isolated
spiral galaxies experience similar chemical evolutionary histories.
We have investigated the evolution and SFHs of M33 and NGC\,300 in
previous work by using the simple chemical evolution model \citep{kang12, kang16}.
Thus, before making a comparative study of these three galaxies, we
should use the simple chemical evolution model to explore the evolution
and SFH of NGC\,2403 first. As a result,
the outline of this paper is organized as follows.
The observational constraints of NGC\,2403 are presented in Section 2.
The main ingredients of our model are described in Section 3.
In Section 4, we focus on investigating the evolution and SFH of
NGC\,2403 in spatial and temporal detail. In section 5,
we compare the SFH of NGC\,2403 with those of NGC\,300 and M33, to
explore the role of environment played in producing their differences
and whether NGC\,2403 and NGC\,300 follow similar SFHs or not.
Section 6 presents a summary of our results.

\section{Observational data}
\label{sect:Obs}

\begin{table}
\caption{Global observational constraints for the disc of NGC\,2403.}
\label{Tab:obs2}
\begin{center}
\begin{tabular}{lll}
\hline
\hline
Property             &      Value                &    Refs.      \\
\hline
Stellar mass         & $\sim5.01\,\times10^{9}~\rm M_{\odot}$             & 1\\
H{\sc i} mass        & $\sim(2.52-5.37)\,\times10^{9}~\rm M_{\odot}$      &1, 2, 3, 4, 5, 6 \\
H$_2$ mass           & $\sim(1.99-7.24)\,\times10^{7}~\rm M_{\odot}$      & 1, 2, 3\\
Gas fraction                                 & $\sim0.336-0.521$        & this paper \\
${\rm 12+log(O/H)_{R_{\rm e}}}$      & $\sim8.33-8.81$            & 2, 7, 8 \\
Total SFR            & $\sim0.382-1.3\,M_{\odot }\rm \,yr^{-1}$           & 1, 3, 4, 9 \\
sSFR                & $\sim\,-10.249\,\pm\,0.025\,\rm yr^{-1}$                 &  10       \\
\hline
\end{tabular}\\[1mm]
\end{center}
Refs: (1) \citet{leroy08}; (2) \citet{garnett02}; (3) \citet{kennicutt03};
(4) \citet{Thilker2007}; (5) \citet{Wiegert14}; (6) \citet{Blok2014};
(7) \citet{moustakas10}; (8) \citet{pilyugin14_1}; (9) \citet{K&K13};
(10) \citet{MM07}
\end{table}

Since a successful galaxy chemical evolution model, especially one
involving free parameters, should reproduce as many observational
constraints as possible. The observed present-day cold
gas, star-formation rate (SFR) and metallicity are crucial
constraints on the galactic evolution model.
Therefore, in this Section, we summarize the current available
observations for  the disc of NGC\,2403, including the radial
distribution and global constraints.

\subsection{Radial distribution of H{\sc i}, SFR and metallicity}

\citet{leroy08} and \citet{Schruba2011} derived the radial distribution
of H{\sc i} gas mass surface density $\Sigma_{\rm HI}(r,t)$ for NGC\,2403
by using Very Large Array (VLA) maps of the 21\,cm line, which is carried
out by the National Radio Astronomy Observatory (NRAO).
\citet{Blok2014} presented deep H{\sc i} observations of NGC\,2403
obtained with the Green Bank Telescope of NRAO.

The SFR surface density, $\Sigma_{\rm SFR}$, was obtained from combinations
of far-ultraviolet (FUV) with $24\,\rm \mu m$ maps \citep{leroy08} and
$\rm H\alpha$ with $24\,\rm \mu m$ maps \citep{Schruba2011}.
\citet{williams13a} estimated the recent radial profiles of $\Sigma_{\rm SFR}$
from the resolved stars. The $\lambda22\,\rm cm$ radio-continuum
emission was used to derive $\Sigma_{\rm SFR}$ by \citet{heesen14}.

Oxygen is the most abundant heavy element in the Universe
\citep{korotin14, zahid14}, and it is always using as a proxy for
the production of all heavy elements in galaxies \citep{zahid14}.
Furthermore, in the research of galaxies, metallicity is defined
as the amount of oxygen \emph{relative} to hydrogen, ${\rm 12+log(O/H)}$.
Searle (1971) was the first to find that disc galaxies might possess
radial metallicity gradient. Subsequently, a number of authors confirmed
his interpretation that the gas-phase metallicity decreases with radius
\citep[e.g., ][]{McCall1985, zaritsky94, RS08, bresolin09, moustakas10, pilyugin14_1}.
Importantly, the observed abundance gradient is one of the important
constraints on the galaxy chemical evolution model \citep{bp00, m&d05, chang12,
kang12, kang16}. The radial distribution of gas-phase oxygen abundance
(${\rm 12+log(O/H)}$) along the disc of NGC\,2403 has been derived by several
authors \citep{McCall1985, zaritsky94, Garnett1997, Garnett1999,
van_Zee1998, Bresolin1999, moustakas10, Berg2013, pilyugin14_1}, and those
adopted to constrain our model are including ${\rm 12+log(O/H)}$ from
\citet{Berg2013}, \citet{pilyugin14_1} and \citet[][e.g., using the
empirical calibrated method in \citet{PT05}]{moustakas10}.

All the above mentioned radial profiles of H{\sc i}, SFR and
${\rm 12+log(O/H)}$ will be used in Figure \ref{Fig:result} to constrain
the model for selecting the best-fitting model for NGC\,2403.

\subsection{Global properties of H{\sc i}, SFR and metallicity}

NGC\,2403 is a H{\sc i}-dominated, low-mass spiral galaxy, and
its H{\sc i} gas disc extends beyond its optical disc.
The atomic hydrogen mass of NGC\,2403 has been calculated to be
$M_{\rm HI}\,\sim\,(2.189-5.37)\times10^{9}\,\rm M_{\odot}$
\citep{garnett02, kennicutt03, Thilker2007, leroy08, Wiegert14}.
Its molecular gas mass $M_{\rm H_{2}}$ is
$M_{\rm H_{2}}\,\sim\,(1.99-7.24)\times10^{7}\,\rm M_{\odot}$
\citep{garnett02, kennicutt03, leroy08}.
\citet{leroy08} estimated the total stellar mass and disc
scale-length of NGC\,2403 from \emph{Spitzer} $3.6\,\rm \mu m$ maps,
and the corresponding values are $M_{\rm *}\,=\,5.0\times10^{9}\rm M_{\odot}$
and $r_{\rm d}\,=\,1.6\,\rm kpc$, respectively. Therefore, we can
easily get the gas fraction $f_{\rm gas}$, which is defined as
$f_{\rm gas}\,=\,\frac{M_{\rm HI}+M_{\rm H_{2}}}{M_{\rm HI}+M_{\rm H_{2}}+M_{\rm *}}$,
and its value is $\sim\,0.336-0.521$.

Using different tracers, the current total SFR for the NGC\,2403 disc
has been measured by several groups,
$1.3\,\rm M_{\odot}\,yr^{-1}$ from $\rm H\alpha$ emission by \citet{kennicutt03},
$0.382\,\rm M_{\odot}\,yr^{-1}$ from a combination of FUV with $24\,\rm \mu m$
maps by \citet{leroy08}, $0.692\,\rm M_{\odot}\,yr^{-1}$ from $\rm H\alpha$ emission
and $0.813\,\rm M_{\odot}\,yr^{-1}$ from FUV luminosity calibrated by \citet{K&K13}, and
$0.912\,\rm M_{\odot}\,yr^{-1}$ derived jointly from TIR and UV luminosities
by \citet{Thilker2007}. \citet{MM07} used FUV$-K$ color to calculate specific SFR
(sSFR), and derived that the value of sSFR for NGC\,2403 is $-10.249\,\pm\,0.025\,\rm yr^{-1}$.
sSFR is defined as $\rm sSFR\,=\,SFR/M_{\ast}$, which represents the ratio of young
to old stars and shows what fraction of total star formation have been occurred
recently.

The gas-phase metallicity of a galaxy is usually represented by the value
of oxygen abundance at the effective radius of the disc,
${\rm 12+log(O/H)_{R_{\rm e}}}$ \citep{zaritsky94, garnett02, sanchez13}.
$R_{\rm e}$ is defined as the radius at which the integrated
flux is half of the total one, and it is equal to 1.685 times the
radial scale-length $r_{\rm d}$ of the disc. The oxygen abundance for the disc
of NGC\,2403 has been measured using different calibrated methods
\citep{zaritsky94, garnett02, moustakas10, pilyugin14_1}.
A summery of the global properties for the disc of NGC\,2403 is
displayed in Table \ref{Tab:obs2}, which will be used to constrain
the model in Section \ref{sect:result}.

\section{The Model}
\label{sec:model} 

In this section, we briefly introduce the basic assumptions and main
ingredients of the model. We assume that the NGC\,2403 disc is
progressively built up by continuous infall of primordial
gas ($X\,=\,0.7571, Y_{\rm p}\,=\,0.2429, Z\,=\,0$) from its halo,
and it is composed of a set of independently evolved concentric
rings with the width 500\,pc.
Infalls of primordial gas, star formation, metal production via stellar
evolution, stellar mass return, and outflows of metal enriched gas are
considered in our model, both instantaneous recycling assumption (IRA)
and instantaneous mixing of the ISM with stellar ejecta are also assumed
in our model. For simplicity, neither radial gas flows nor stellar migration
is taken into account in our model. We adopt a standard flat cosmology with
$\Omega_{\rm m}\,=\,0.3$, $\Omega_{\Lambda}\,=\,0.7$, and the Hubble constant
$h\,=\,H_{0}/100{\rm \,km\,s^{-1}\,Mpc^{-1}}$ in this work, and the adopted
solar metallicity is $12+{\rm log(O/H)}=8.69$ \citep{asplund09}.

Following \citet{kang12, kang16}, the classical set of integro-differential
equations from \citet{Tinsley80} are adopted to express the chemical evolution
in each ring of the disc for NGC\,2403. Since the return fraction $R$ and
the nucleosynthesis yield $y$ in the chemical evolution model mainly depend on
the adopted stellar initial mass function (IMF). We obtain $R=0.43$ and
$y=1\,{\rm Z}_{\odot}=0.02$ after taking stellar IMF from \citet{chabrier03}.

\citet{Bigiel2014} identified clear inflow signatures from the study of
H{\sc i} velocity fields for NGC\,2403.
The cold gas infall rate ($f_{\rm in}(r,t)$, in units of
$\rm{M_{\odot}}\,{pc}^{-2}\,{Gyr}^{-1}$), as a function of space and time,
is adopted as that in \citet{kang12} and \citet{kang16}:
\begin{equation}
f_{\rm{in}}(r,t)=A(r)\cdot t\cdot e^{-t/\tau},
\label{eq:infall rate}
\end{equation}
where $\tau$ is the gas infall timescale, and it is a free parameter
in our model. The $A(r)$ are a set of separate quantities
constrained by the present-day stellar mass surface density
$\Sigma_*(r,t_{\rm g})$, and $t_{\rm g}$ is the cosmic age.
we set $t_{\rm g}=13.5\rm\,Gyr$ according to the standard
flat cosmology.
That is, $A(r)$ are iteratively calculated by requiring the
model predicted $\Sigma_*(r,t_{\rm g})$ equal to its observed value
\citep{chang12, kang12, kang16}. The present-day stellar mass surface
density of NGC\,2403 is well described by a simple exponential profile
\citep{Cepa1988, barker2012, williams13a},
\begin{equation}
\Sigma_*(r,t_{\rm g})=\Sigma_*(0,t_{\rm g}){\rm exp}(-r/r_{\rm d}),
\label{eq:stellar_distrbution}
\end{equation}
where $r_{\rm d}$ is the present-day value of disc scale-length.
$\Sigma_*(0,t_{\rm g})$ is the present-day central stellar mass
surface density, and it can be easily obtained from
$\Sigma_*(0,t_{\rm g})=M_{*}/2\pi r_{\rm d}^{2}$. Here, we adopt the
stellar mass and disc scale-length of NGC\,2403
as $M_{*}\,=\,5.0\times10^{9}~\rm M_{\odot}$
and $r_{\rm d}\,=\,1.6\,\rm kpc$, which are obtained from the $3.6\,\mu m$
luminosity \citep{leroy08}.

The SFR surface density, $\Psi(r,t)$, describes the total mass of newly born
stars in unit time and area. At each radius $r$ and time $t$, $\Psi(r,t)$
depends on the local molecular gas surface density $\Sigma_{\rm H_2}(r,t)$
\citep{leroy08, leroy13, chang12, kang12, kang16, Kubryk2015},
and $\Psi(r,t)$ is linearly proportional to $\Sigma_{\rm H_2}(r,t)$:
\begin{equation}
\Psi(r,t)=\Sigma_{\rm{H_2}}(r,t)/t_{\rm dep},
\label{eq:h2sfr}
\end{equation}
where $t_{\rm dep}$ is the molecular gas depletion time, and its value
is adopted as $t_{\rm dep}\,=\,1.9\,\rm Gyr$ in this work \citep{leroy08,leroy13}.
The reader is referred to \citet{kang12, kang16} for a more indepth description
for the calculation of the molecular hydrogen to atomic hydrogen gas surface
density in a galaxy disc, as well as the reason that we adopt the $\rm H_2$
correlated star formation law.

In general, low-mass galaxies have low escape velocities due to their
shallower
gravitational potentials, which make them more susceptible to losing
their ISM through supernova feedback \citep{H&T1995, Thuan99,tremonti04}.
In addition, galaxies with rotation speed
$V_{\rm rot}\leq\,100-150\,{\rm km s^{-1}}$ may expel a large part of
their supernova ejecta to the circumgalactic medium \citep{garnett02}.
NGC\,2403 is a fairly low-mass disc galaxy with stellar mass
$M_{\ast}\,=\,5.01\times10^{9.7}\,\rm {M_{\odot}}$ \citep{leroy08} and
with a rotation speed about $V_{\rm rot}\approx 136\,{\rm km s^{-1}}$
\citep{garnett02}, thus the gas-outflow process has a significant
influence on the chemical enrichment during its evolution history.

The outflowing gas is assumed to not fall again to the disc, and it
has the same metallicity as the ISM when the outflow process occurred
\citep{chang10, kang12, kang16, ho15}. Following the method of
\citet{recchi08}, the gas outflow rate
$f_{\rm out}(r,t)$ (in units of $\rm{M_{\odot}}\,{pc}^{-2}\,{Gyr}^{-1}$)
is assumed to be proportional to $\Psi(r,t)$, that is:
\begin{equation}
  f_{\rm out}(r,t)=b_{\rm out}\Psi(r,t)
\label{eq:outflow}
\end{equation}
where $b_{\rm out}$ is the outflow efficiency, and it is the other free
parameter in our model.

In summary, there are two free parameters, the infall time-scale
$\tau$ and the outflow efficiency $b_{\rm out}$, left in our model.
Moreover, there exists degeneracy between $y$ and $b_{\rm out}$ in that
the model adopting a higher $y$ need a larger $b_{\rm out}$ to reproduce
the observed metallicity profile.
Thanks to the fact that the reasonable range of $y$ is small compared with the
large possible rang of $b_{\rm out}$, we can constrain $b_{\rm out}$ using
the observed abundance distribution.

\section{Results}
\label{sect:result}

\begin{figure*}
   \centering
   \includegraphics[angle=0,scale=0.9]{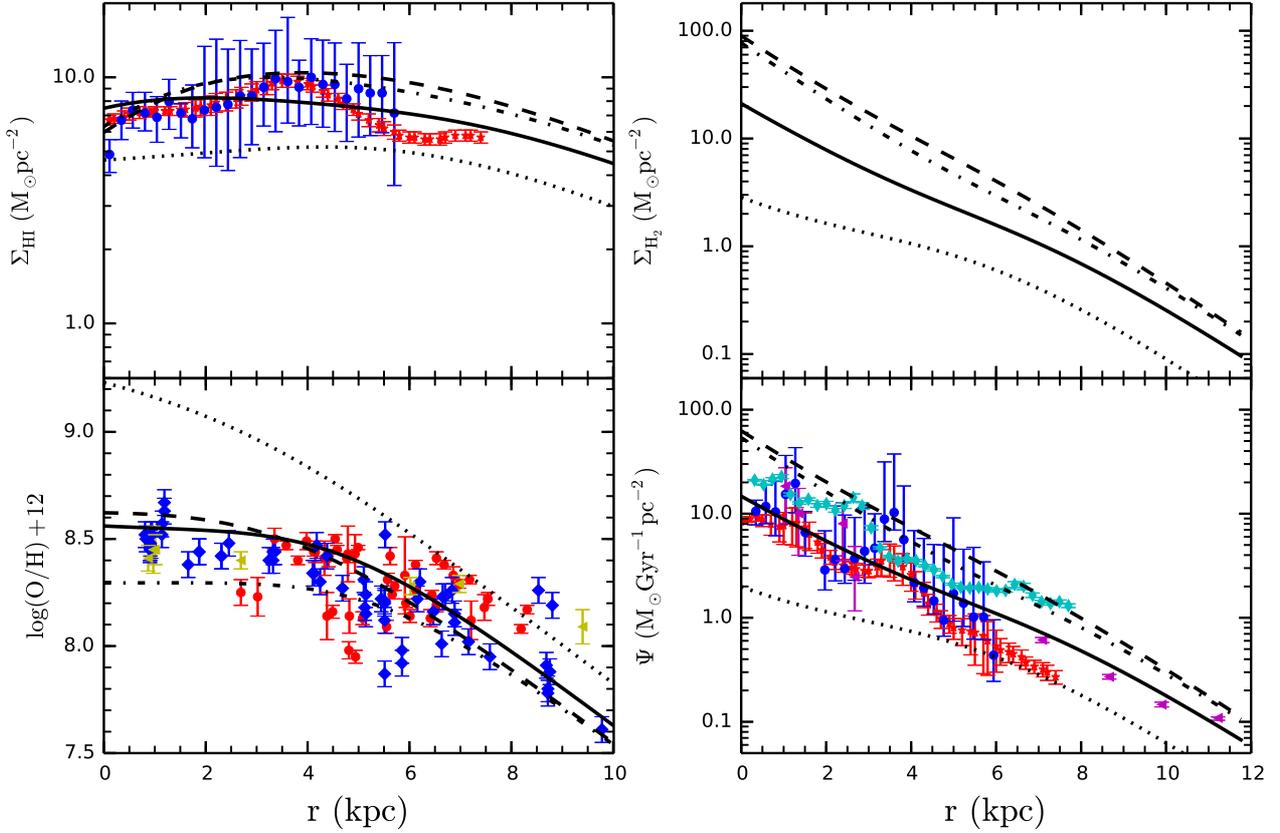}
   \caption{Comparisons of the radial profiles between the
   model predictions and the observational data. Different
   line types are corresponding to various parameter groups:
   dotted lines $(\tau, b_{\rm out})=(0.1\,{\rm Gyr}, 0)$, dashed
   lines $(\tau, b_{\rm out})=(15\,{\rm Gyr}, 0)$, dot-dashed lines
   $(\tau, b_{\rm out})=(15\,{\rm Gyr}, 1)$, solid lines
   $(\tau, b_{\rm out})=(0.2r/{\rm r_{d}}+3.2\,{\rm Gyr}, 0.6)$.
   The radial profiles of H{\sc i} mass surface density and oxygen abundance
   are separately shown in the top and bottom panels of the left-hand side;
   On the right-hand side, the radial profiles of H2 mass and SFR surface
   density are displayed in the top and bottom panels, respectively.
   H{\sc i} data from \citet{leroy08} are shown by red filled asterisks,
   while those from \citet{Schruba2011} are displayed by blue filled circles.
   The observed oxygen abundance taken from \citet{moustakas10}, \citet{Berg2013}
   and \citet{pilyugin14_1} are plotted by red filled circles, magenta filled
   triangles and blue filled diamonds, respectively.
   SFR data obtained from \citet{leroy08}, \citet{Schruba2011}, \citet{williams13a}
   and \citet{heesen14} are separately denoted as red filled asterisks,
   blue filled circles, magenta filled triangles and cyan filled diamonds.
   It should be pointed out that the SFR data from \citet{williams13a} are the
   recent SFR as a function of radius.
   }
   \label{Fig:result}
\end{figure*}

\begin{figure}
   \includegraphics[angle=0,scale=0.55]{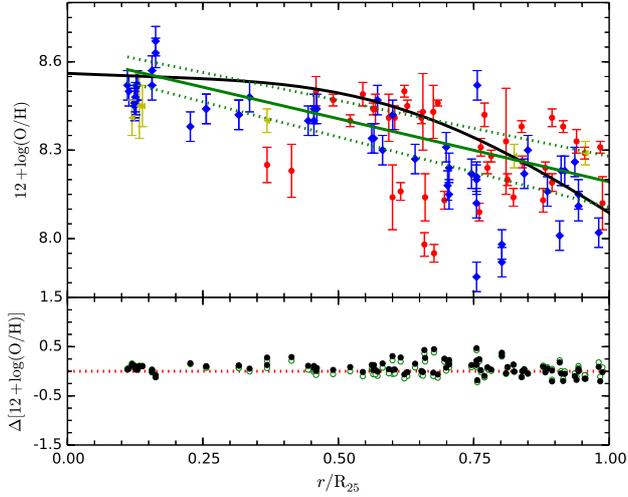}
   \caption{\emph{Upper panel}: comparisons of the present-day radial profiles of
   $\rm 12+log(O/H)$ predicted by our best-fitting model (black solid line) with
   the observational data when the galactocentric are normalized to ${R_{\rm 25}}$, the
   symbols are the same as that in the lower panel (left column) of Figure \ref{Fig:result}.
   The green solid line represents the linear best fit to the observed data, and the
   green dotted lines are the $1\sigma$ uncertainty.
   \emph{Lower panel}: the black filled circles show the deviations of the observations
   from our best-fitting model predictions, derived by the solid line minus the observational
   data in the above panel, while the green open circles show the ones obtained from the
   linear best fit minus the observational data in the above panel.
   }
\label{Fig:Z_dev}
\end{figure}

\begin{table*}
\caption{The best-fitting model predictions of the total quantities of NGC\,2403.}
\label{Tab:predictions}
\begin{center}
\begin{tabular}{llllllll}
\hline
\hline
$M_{\rm K}$\,(mag) &  $M_{\rm B}$\,(mag)  &  $M_{\rm {H_2}}$\,$(10^{9}\,{\rm M_{\odot}})$
&  $M_{\rm {H_I}}$\,$(10^{9}\,{\rm M_{\odot}})$    & $f_{\rm gas}$
&  SFR\,(${\rm M_{\odot}\,yr^{-1}}$) &  sSFR\,(${\rm yr^{-1}}$)   & ${\rm 12+log(O/H)}_{R_{\rm e}}$ \\
\hline
-21.94      &  -18.73   &   0.797   &  3.28   &  0.469  &  0.418   &  -10.042 &  8.529  \\
\hline
\end{tabular}\\
\end{center}
\end{table*}

As has been mentioned in Section \ref{sec:model} that there are two free parameters,
the gas infall timescale $\tau$ and the outflow efficiency $b_{\rm out}$,
in the model, here we investigate the free parameters on the model results
through comparing the model predictions with observations.
Figure \ref{Fig:result} demonstrates the comparisons between the model predicted
radial profiles and the observations. The dotted line, dashed line and
dot-dashed line are corresponding to three limiting cases of
$(\tau, b_{\rm out})=(0.1\,{\rm Gyr}, 0)$, $(\tau, b_{\rm out})=(15.0\,{\rm Gyr}, 0)$
and $(\tau, b_{\rm out})=(15.0\,{\rm Gyr}, 1.0)$, respectively.
Comparisons among dotted line, dashed line and dot-dashed line show that
the model predictions are very sensitive to the adopted infall timescale
$\tau$, while the outflow efficiency $b_{\rm out}$ mainly influence the
shape of metallicity. This is mainly due to the fact that the infall
timescale $\tau$ mainly determines the gas supply during the whole
evolutionary history of a galaxy, while the outflow efficiency $b_{\rm out}$
takes a fraction of metals away from the disc.

It can be seen from Figure \ref{Fig:result} that almost all the observational
constraints locate between the dotted line and the dot-dashed line, which implies
that we can certainly find a model that can reproduce the main observational
data along the disc of NGC\,2403. Figure \ref{Fig:result} also shows that
the metallicity in the central region is high than that in the outmost region,
and the observed results certify that the disc of NGC\,2403 follows an
inside-out formation scenario \citep{MM07}. Like the method in our previous work
\citep{kang12,kang16}, we assume the form of infall timescale is
$\tau(r)\,=\,a\times r/{\rm r_{d}}+b$ \citep{chiosi1980, matteucci89},
where a and b are the coefficients for $\tau(r)$. Including another
free parameter $b_{\rm out}$, there are three free parameters (a, b and
$b_{\rm out}$) in our model that we should determine.

In order to select the best-fitting model for NGC\,2403, we should
first search for the best combination of free parameters a, b and
$b_{\rm out}$. Thus, we use the classical $\chi^{2}$ technique
through comparing the model predicted profiles with the corresponding
observational data, such as the radial profiles of H{\sc i}, SFR and
$\rm 12+log(O/H)$ \citep[like the method adopted in ][]{kang16}. The
boundary conditions of a, b and $b_{\rm out}$ for NGC\,2403
are separately assumed to be $0\leq\,a\,\leq1.5$, $1.0\leq\,b\leq\,5.0$
and $0.1\leq\,<b_{\rm out}\leq\,0.9$. In practice, we calculate the
value of $\chi^{2}$ by comparing our model predictions with the observed
data, i.e., the combination of the radial profiles of H{\sc i}, SFR and
$\rm 12+log(O/H)$. For each pair of a and b, we vary the value
of $b_{\rm out}$ to search for the minimum value of $\chi^{2}$. The minimum
value of $\chi^{2}$ is corresponding to the best combination of a, b and
$b_{\rm out}$, and we obtain (a, b, $b_{\rm out}$)= (0.2, 3.2, 0.6). That is,
$(\tau, b_{\rm out})=(0.2r/{\rm r_{d}}+3.2\,{\rm Gyr}, 0.6)$
is defined as the best-fitting model, and its results are shown as
solid lines in Fig. \ref{Fig:result}. It can be found that the solid lines
almost reproduce all the observed data, which means that
this parameter group may reasonably describe the crucial
ingredients of the main physical processes that regulate the formation
and evolution of NGC\,2403.

The metallicity provides an important indicator of the evolutionary
history of a galaxy, therefore we use $\rm 12+log(O/H)$ to further state the
best-fitting model can nicely describe the formation and evolution of
NGC\,2403. Comparisons of the best-fitting model predicted present-day
$\rm 12+log(O/H)$ distributions (black solid line) with the observational data (points)
are displayed in the upper panel of Figure \ref{Fig:Z_dev} when the galactocentric are
normalized to ${R_{\rm 25}}$. The deviations of the observed $\rm 12+log(O/H)$ points
from our best-fitting model predicted $\rm 12+log(O/H)$ are shown as the black filled
circles in the lower panel of Figure \ref{Fig:Z_dev}, which are obtained from the
best-fitting model predictions minus the observational data.
We also find the linear best fit to the observed data within the optical radius.
The linear best fit and the $1\sigma$ uncertainty are also shown in the upper panel
of Figure \ref{Fig:Z_dev} by the green solid line and green dotted lines,
respectively. The deviations of the observed $\rm 12+log(O/H)$ points from the
corresponding linear best fit are denoted as the green open circles in the lower panel
of Figure \ref{Fig:Z_dev}. The lower panel of Figure \ref{Fig:Z_dev} reveals
that the black solid circles almost coincide with the green open circles. The mean
values of deviations for the black filled circles and the green open circles are
0.12 and 0.11, respectively.

\begin{figure}
   \includegraphics[angle=0,scale=0.55]{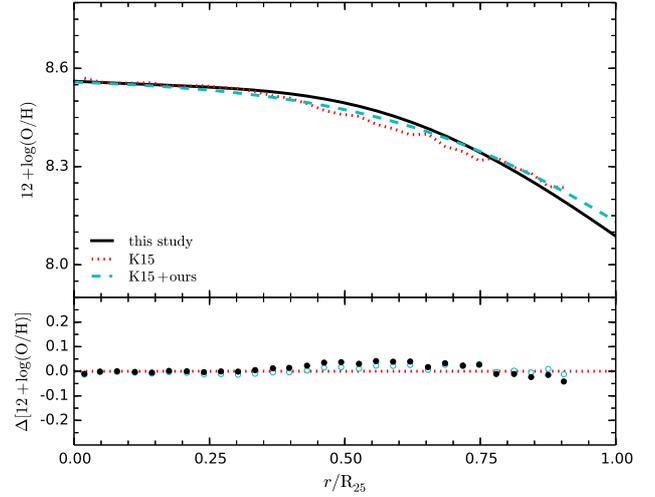}
   \caption{\emph{Upper panel}: comparisons of the present-day radial profiles of
   $\rm 12+log(O/H)$ predicted by our best-fitting model (black solid line) with
   that of \citet{Kudritzki2015} (red dotted line) when the radius is measured in
   ${R_{\rm 25}}$. The cyan dashed line represents the radial profile of $\rm 12+log(O/H)$
   derived by replacing the observed ratio of stellar to gas mass surface density in
   equation \ref{eq:z1} of \citet{Kudritzki2015} with our best-fitting model predicted ones.
   \emph{Lower panel}: the black filled circles represent the deviations of $\rm 12+log(O/H)$
   derived by the solid line minus the red dotted line in the above panel,
   while the cyan open circles show the ones obtained from the cyan dashed line
   minus the red dotted line in the above panel.
   }
\label{Fig:me_K15}
\end{figure}

Recently, \citet{Kudritzki2015} derived the present-day radial
profile of $\rm 12+log(O/H)$ for NGC\,2403 through
${\rm 12+log(O/H)}(r, t_{\rm g})\,=\,{\rm log10}(Z(r, t_{\rm g})/(16.0\times0.710))$, and
$Z(r, t_{\rm g})$ is obtained from the following equation,
\begin{equation}\label{eq:z1}
Z(r, t_{\rm g})\,=\,{y_O \over \Lambda}\left\{1-\left[1+(1+\alpha){\Sigma_{*}(r, t_{\rm g})\over \Sigma_{\rm gas}(r, t_{\rm g})}\right]^ {-w}\right\},
\end{equation}
where $w\,=\,{\Lambda \over (1-R)(1+\alpha)}$ and $\alpha\,=\,{\eta- \Lambda \over 1-R}$,
$\eta$ and $\Lambda$ are the mass-loss and the mass-gain loading factors
respectively, and $\eta\,=\,0.81$ and $\Lambda\,=\,0.71$ for NGC\,2403. The
return fraction $R\,=\,0.4$ and the oxygen yield $y_O\,=\,0.00313$ are adopted.
$\Sigma_{*}(r, t_{\rm g})$ and $\Sigma_{\rm gas}(r, t_{\rm g})$ are the observed
present-day stellar mass and total gas mass surface density from \citet{Schruba2011},
and the total gas mass surface density is defined as
$\Sigma_{\rm gas}(r, t_{\rm g})\,=\,\Sigma_{\rm{H_2}}(r,t_{\rm g})\,+\,\Sigma_{\rm{HI}}(r,t_{\rm g})$
(More information about the derivation of the models and equation
\ref{eq:z1} are in \citet{Kudritzki2015}). Here, we compare the radial profile
of $\rm 12+log(O/H)$ predicted by our best-fitting model with that
of \citet{Kudritzki2015}, which are plotted in Fig. \ref{Fig:me_K15} by
black solid line and red dotted line, respectively. It can be found that
our results are basically agreement with their results.
Using the radial profile of $\rm 12+log(O/H)$ along the disc of NGC\,2403
predicted by our model and that predicted by the recent model of \citet{Kudritzki2015}
to calculate the deviations of $\rm 12+log(O/H)$ for NGC\,2403, which are denoted as
black filled circles in the lower panel of Fig.\ref{Fig:me_K15}, we further demonstrate
there is no significant difference between the two predictions.

In order to further certify our results, we also use our best-fitting
model predicted
$\Sigma_{*}(r, t_{\rm g})$ and $\Sigma_{\rm gas}(r, t_{\rm g})$ to replace
the observed $\Sigma_{*}(r, t_{\rm g})$ and $\Sigma_{\rm gas}(r, t_{\rm g})$
in equation \ref{eq:z1}, and other parameters in the equation \ref{eq:z1} are
the same as those of \citet{Kudritzki2015}. The result is displayed as cyan
dashed line in Fig. \ref{Fig:me_K15}. It can be seen that the cyan dashed line
and the red dotted line are nearly the same, and their corresponding deviations
of $\rm 12+log(O/H)$ are displayed as cyan open circles in the lower
panel of Fig.\ref{Fig:me_K15}. Most important of all, the mean value
of deviation for the results of \citet{Kudritzki2015} (red dotted line in the
upper panel of Fig.\ref{Fig:me_K15}) from the observed $\rm 12+log(O/H)$ data
(points in the upper panel of Fig.\ref{Fig:Z_dev}) is $0.11$, and the mean
deviation value for K$15+$ours (cyan dashed line in the upper panel of
Fig.\ref{Fig:me_K15}) from the observed $\rm 12+log(O/H)$ data (points in the
upper panel of Fig.\ref{Fig:Z_dev}) is also $0.11$.
This information reinforces our results that our model
predicted $\Sigma_{*}(r, t_{\rm g})$ and $\Sigma_{\rm gas}(r, t_{\rm g})$ can
also nicely reproduce the observed values from \citet{Schruba2011}.

In Table \ref{Tab:predictions}, we also display the best-fitting model predicted
present-day total quantities of NGC\,2403, such as the absolute $K-$ and
$B-$ band magnitude $M_{\rm K}$ and $M_{\rm B}$, H{\sc i} mass, gas
fraction $f_{\rm gas}$, total SFR, sSFR and characteristic
$\rm 12+log(O/H)$ (defined as the oxygen value $\rm 12+log(O/H)$ at
the effective radius $R_{\rm e}$). It can be found that the physical
quantities in Table \ref{Tab:predictions} are basically agreement with
the corresponding observed ones in Table \ref{Tab:obs1} and Table \ref{Tab:obs2}
considering the observed uncertainties.

\section{Comparison with NGC\,300 and M33}

\begin{table*}
\caption[best]{The main input properties and parameters of the best-fitting
models for NGC\,2403, NGC\,300 and M33.}
\begin{center}
\begin{tabular}{llllll}
\hline
Individual                                  &  & &  NGC\,2403     &      NGC\,300     &     M33        \\
\hline
Input physical properties &  &  Total stellar mass ($10^{9}\rm M_{\odot}$)  &   5.0    &    1.928      & 4.0   \\
 &  &  Scale-length $r_{\rm d}$ (kpc)              &  1.6           &        1.29       & 1.4                   \\
Star formation law         &     $\Psi(r,t)\,=\,\Sigma_{\rm H_{2}}(r,t)/t_{\rm dep}$ &   Molecular gas depletion time\,$t_{\rm dep}\,(\rm Gyr)$                     &    1.9           &        1.9        &   0.46     \\
Infall rate  &    $f_{\rm{in}}(r,t)\,\propto\,t\cdot e^{-t/\tau}$          &    Infall time-scale\,$\tau(r)$\,(Gyr)            &  $0.2r/{r_{\rm d}}+3.2$    &$0.35r/{r_{\rm d}}+2.47$   &  $r/{r_{\rm d}}+5.0$  \\
Outflow rate    &   $f_{\rm out}(r,t)\,=\,b_{\rm out}\Psi(r,t)$         &   Outflow efficiency\,$b_{\rm out}$            &    0.6          &         0.9        &  $0.5$         \\
\hline
\end{tabular}
\end{center}
\label{tab:best}
\end{table*}

\begin{figure}
   \includegraphics[angle=0,scale=0.55]{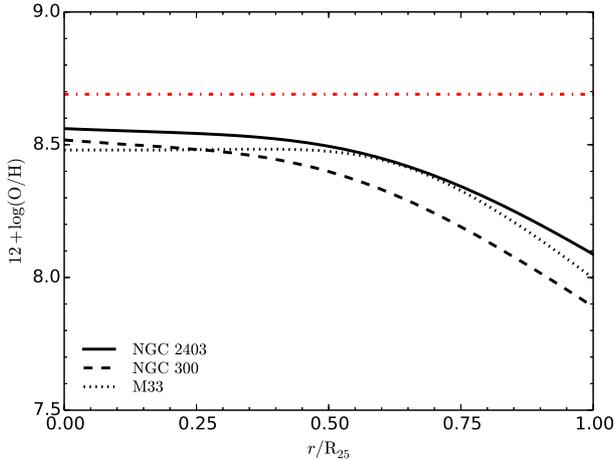}
   \caption{The present-day radial profiles of $\rm 12+log(O/H)$
   predicted by their own best-fitting models when
   the radius is measured in ${R_{\rm 25}}$. Different line types
   represent the radial profile of $\rm 12+log(O/H)$ for NGC\,2403
   (solid line), NGC\,300 (dashed line) and
   M33 (dotted line). The horizontal red dot-dashed line
   denotes the corresponding solar value in \citet{asplund09}.
   }
\label{Fig:metallicity}
\end{figure}

In order to provide a picture of the local environment on the evolution
of galaxies and whether or not the isolated spiral galaxies follow similar
evolution histories, we will compare the SFH of NGC\,2403 with those of nearby
morphological twins NGC\,300 and M33 in this section. Table \ref{tab:best}
summarizes the main input properties and the best-fitting parameters
for them. Physical details of the model description and searching
for the best-fitting models for NGC\,300 and M33 are in \citet{kang16}
and in \citet{kang12}, respectively. Through comparative studying the
metallicity gradient and mean age gradient, along with cosmic evolution
of disc scale-length, stellar mass and SFR for these three galaxies, we
hope to investigate the impact of local environment on the evolution and
SFH of these three galaxies, as well as whether or not the isolated spiral
galaxies experience similar evolution histories.

\subsection{The metallicity gradient}

\citet{Rupke2010} found that the gas-phase oxygen abundance gradients in
interacting disc galaxies are shallower than that in isolated disc
galaxies, and the median metallicity gradient in interacting and isolated galaxies
is $-0.23\,\pm\,0.03\,{\rm dex}/R_{25}$ and $-0.57\,\pm\,0.05\,{\rm dex}/R_{25}$,
respectively. Furthermore, \citet{pilyugin14_1} measured the metallicity gradients of 130
nearby late-type galaxies, and derived the universal metallicity gradients of
$0.32\,\pm\,0.20\,{\rm dex}\,R_{25}^{-1}$ for 104 of their field galaxies
(i.e., excluding mergers and close pairs). \citet{ho15} also
studied the integral field unit (IFU) observations of 49 isolated spiral
galaxies with absolute magnitudes $-22\,<\,M_{B}\,<\,-16$,
and found evidence for a common metallicity gradient
among their galaxies when the slope is expressed in units of the
isophotal radius $R_{\rm 25}$, i.e., $-0.39\,\pm\,0.18\,{\rm dex}\,R_{25}^{-1}$.

The aforementioned information indicates the necessity to explore the
metallicity gradients for the discs of NGC\,2403, NGC\,300 and M33.
The best-fitting model predicted present-day gas-phase oxygen abundance
profiles of NGC\,2403 (solid line), NGC\,300 (dashed line) and M33
(dotted line) are plotted in Fig. \ref{Fig:metallicity}. The best-fitting
model predicted present-day central oxygen abundance are 8.560 for NGC\,2403,
8.518 for NGC\,300 and 8.480 for M33, consistent with the observed central
value of them in \citet{pilyugin14_1}, i.e., $8.48\pm0.02$ for NGC\,2403,
$8.51\pm0.02$ for NGC\,300 and $8.48\pm0.02$ for M33, taking into account
the observed uncertainties and the calibration for determining metallicity.

It can be seen from Fig. \ref{Fig:metallicity} that the metallicity
gradients of NGC\,2403 and NGC\,300 are similar, while the metallicity
gradient of M33 in the inner region ($\leq\, 0.635R_{25}$) is
obviously shallower than that of NGC2403 and NGC\,300, when the metallicity
gradients are expressed in ${\rm dex}\,R_{\rm 25}^{-1}$. The model predicted
present-day metallicity gradients in the central parts (up to their break radii)
are $-0.154{\rm dex}/R_{25}$ for NGC\,2403, $-0.167{\rm dex}/R_{25}$ for
NGC\,300 and $-0.065{\rm dex}/R_{25}$ for M33. This is mainly because the
interactions between M33 and M31 drive large gas flows towards the
central region, carrying less enriched gas from the outskirts of M33 into
its central region, diluting the central region metallicity, and change
its metallicity gradient.

Compared to the inner regions, the model predicted present-day
metallicity gradients in the outer parts (after their break radii) are
similar for them, e.g., $-0.854{\rm dex}/R_{25}$ for NGC\,2403,
$-0.867{\rm dex}/R_{25}$ for NGC\,300 and $-0.993{\rm dex}/R_{25}$ for M33.
Our model predicted present-day metallicity gradients for these three target
galaxies are in contradiction with the recent results of \citet{TSC2016}, who
carried out the deep spectrophotometry of HII regions in NGC\,300 and M33, and
found that the gradients of oxygen abundances (derived from the collisionally
excited lines) in NGC\,2403, NGC\,300 and M33 are close to each other.
Fortunately, our results are in line with the previous observed results of
\citet{pilyugin14_1}, that is, the global gradients of oxygen abundances in M33 is
shallower than that of NGC\,2403 and NGC\,300. Furthermore,
the similar metallicity gradients in both NGC\,2403 and NGC\,300
indicate that they may experience similar chemical evolutionary
histories \citep{PB2000, ho15}.

\begin{figure*}
  \centering
  \includegraphics[angle=0,scale=0.55]{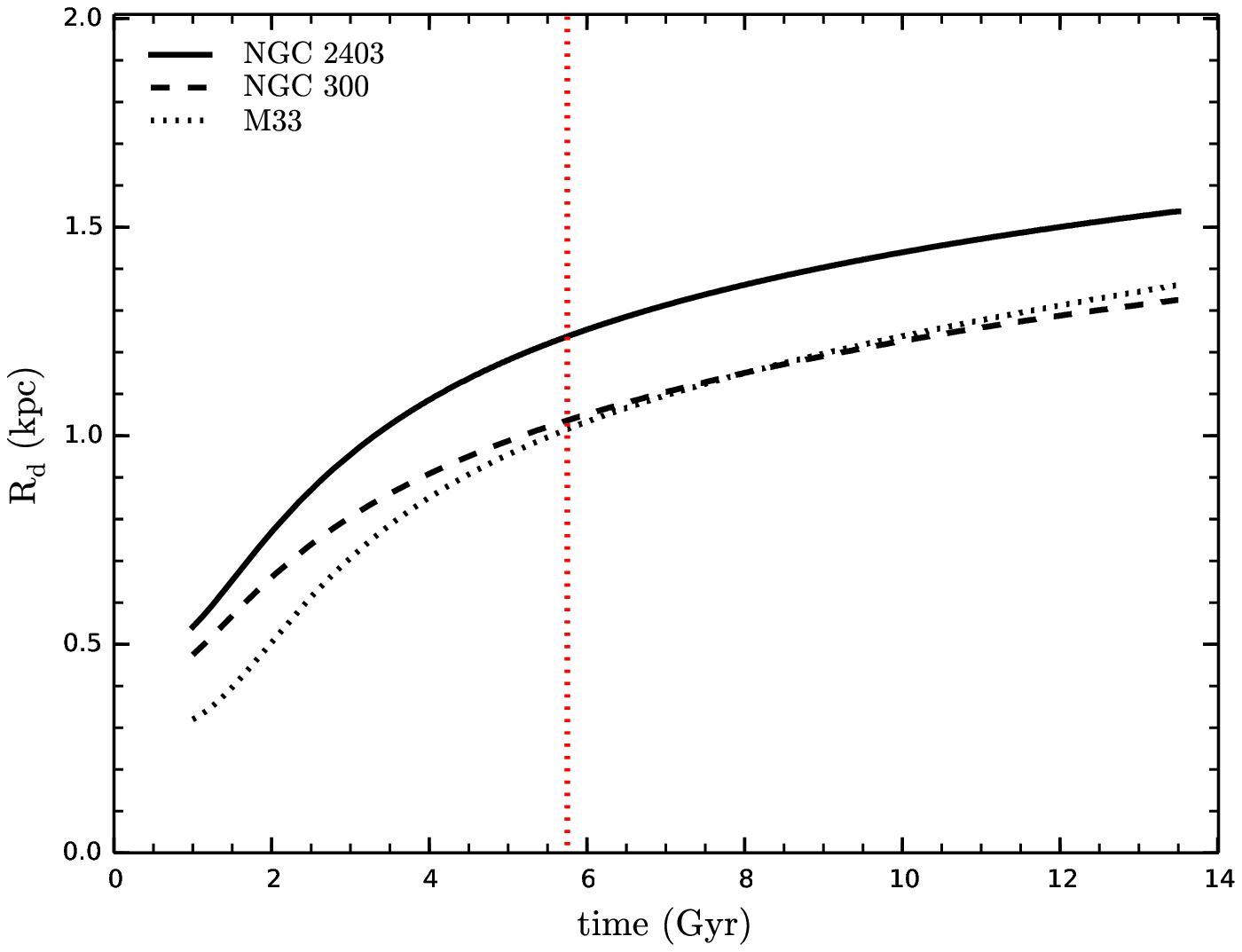}
  \includegraphics[angle=0,scale=0.55]{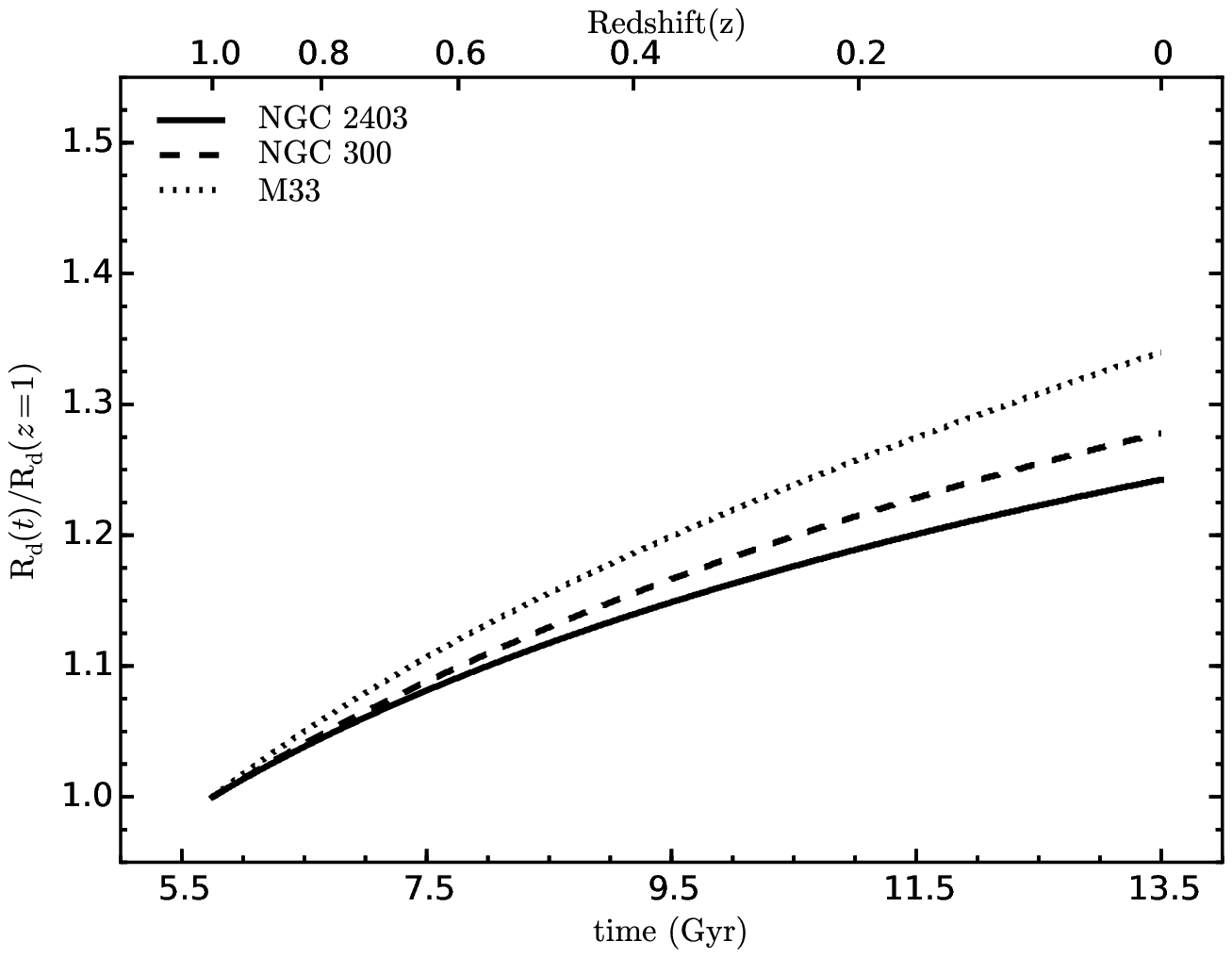}
    \caption{Left panel shows the temporal evolution of the \emph{absolute}
    scale-length $r_{\rm d}$ of the three galaxy discs, and the \emph{relative}
    scale-length growth from $z\,=\,1$ to $z\,=\,0$ is plotted in the right panel.
    Different line types are corresponding to different galaxies: solid line for
    NGC\,2403, dashed line for NGC\,300, and dotted line for M33. The vertical red
    dotted line in the left panel denotes the galaxy evolutionary age at
    $z\,=\,1\,({\rm i.e.,}\,t\,=\,5.75\,\rm Gyr)$.
    }
  \label{Fig:rd_e}
\end{figure*}

\begin{figure}
  \centering
  \includegraphics[angle=0,scale=0.55]{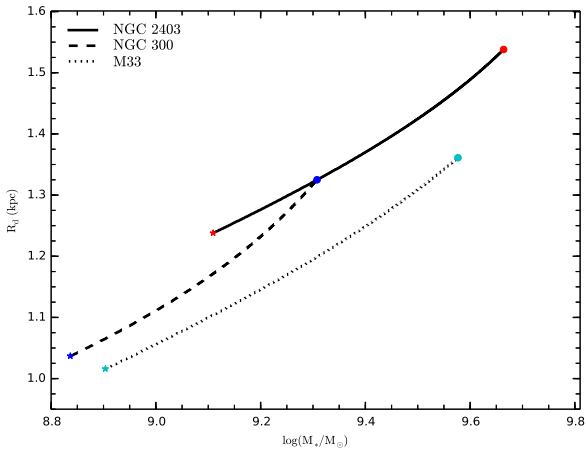}
    \caption{Temporal evolution (from $z\,=\,1$ to $z\,=\,0$) of the
    scale-length and total stellar-mass predicted by our best-fitting
    models for NGC\,2403 (solid line), NGC\,300 (dashed line) and M33
    (dotted line). The $z\,=\,0$ step is marked
    with filled circle, and the $z\,=\,1$ step is marked with asterisk.
    }
  \label{Fig:MS}
\end{figure}

\begin{figure*}
  \centering
  \includegraphics[angle=0,scale=0.55]{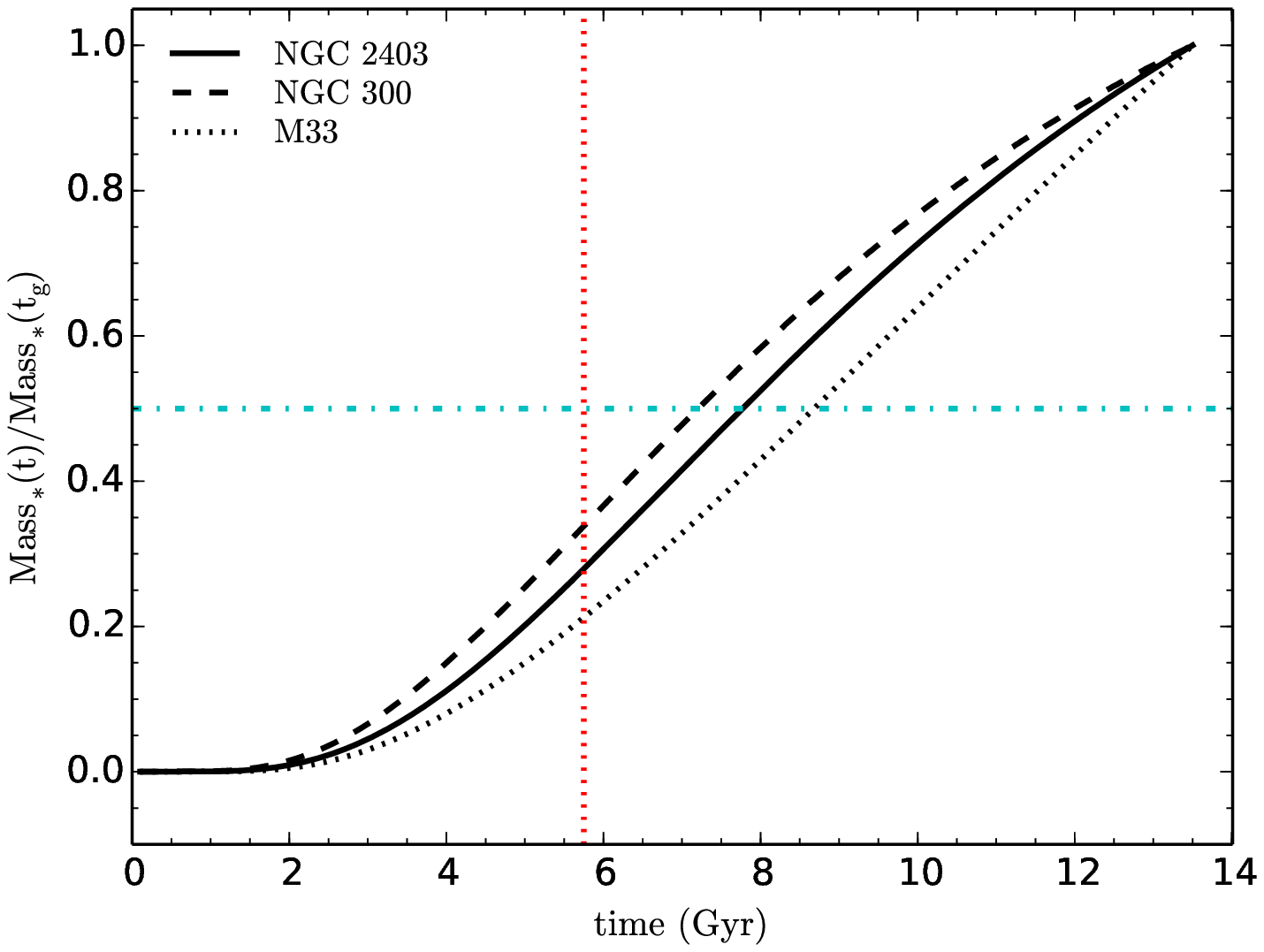}
  \includegraphics[angle=0,scale=0.55]{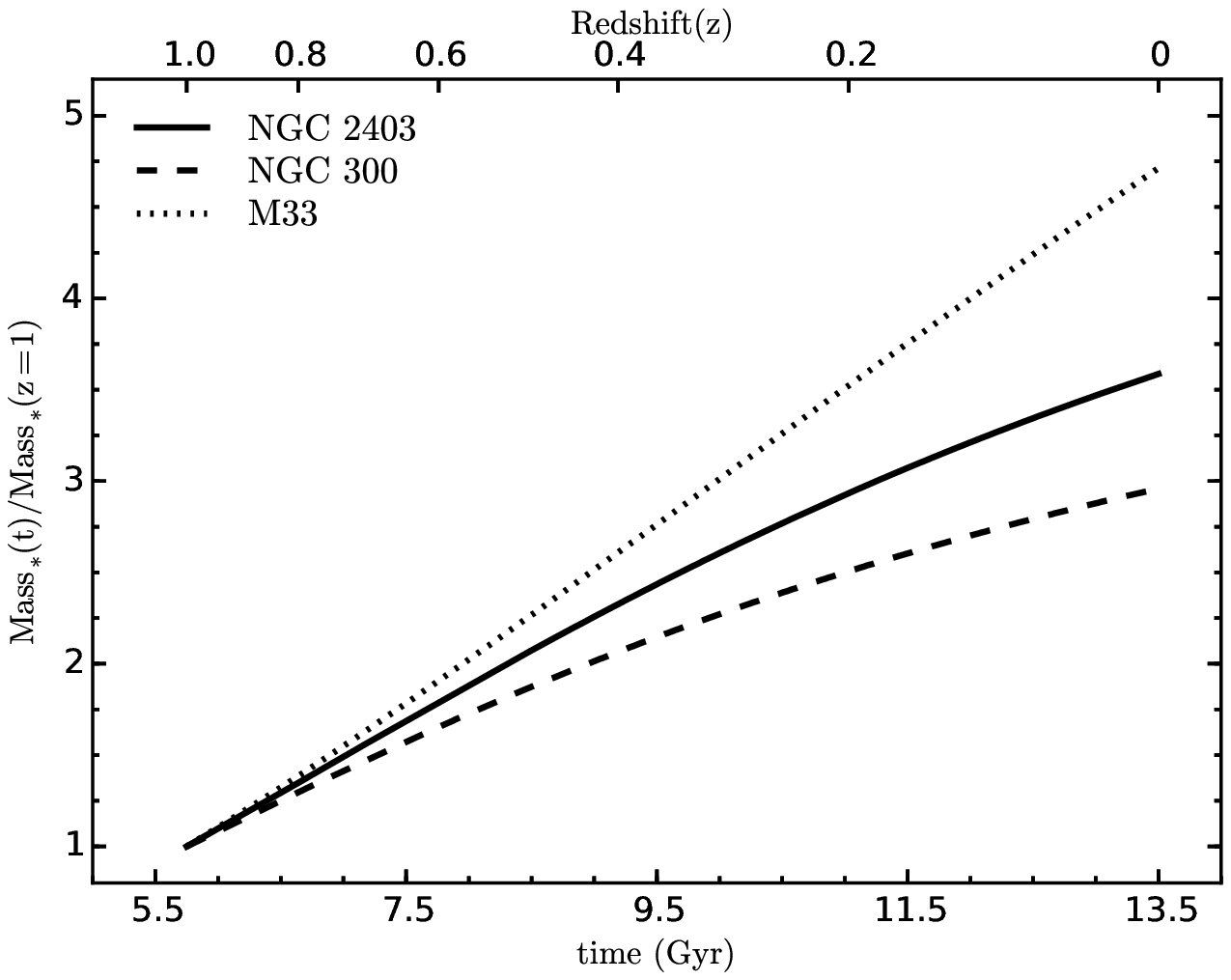}
    \caption{Left panel shows the stellar mass growth histories of
    NGC\,2403 (solid line), NGC\,300 (dashed line) and M33
    (dotted line) predicted by their own best-fitting models. Stellar masses
    are normalized to their present-day values. The horizontal cyan dot-dashed
    line in the panel marks when the stellar mass achieves 50\% of its
    final value, while the vertical red dotted line denotes the galaxy evolutionary
    age at $z\,=\,1\,({\rm i.e., }\,t\,=\,5.75\,\rm Gyr)$. The relative accumulated
    history of stellar mass for them since $z\,=\,1$ is shown in the right panel,
    here normalized to their values at $z\,=\,1$. The line types in the right panel
    are the same as those in the left panel.
    }
  \label{Fig:stellar}
\end{figure*}

\begin{figure*}
   \includegraphics[angle=0,scale=0.55]{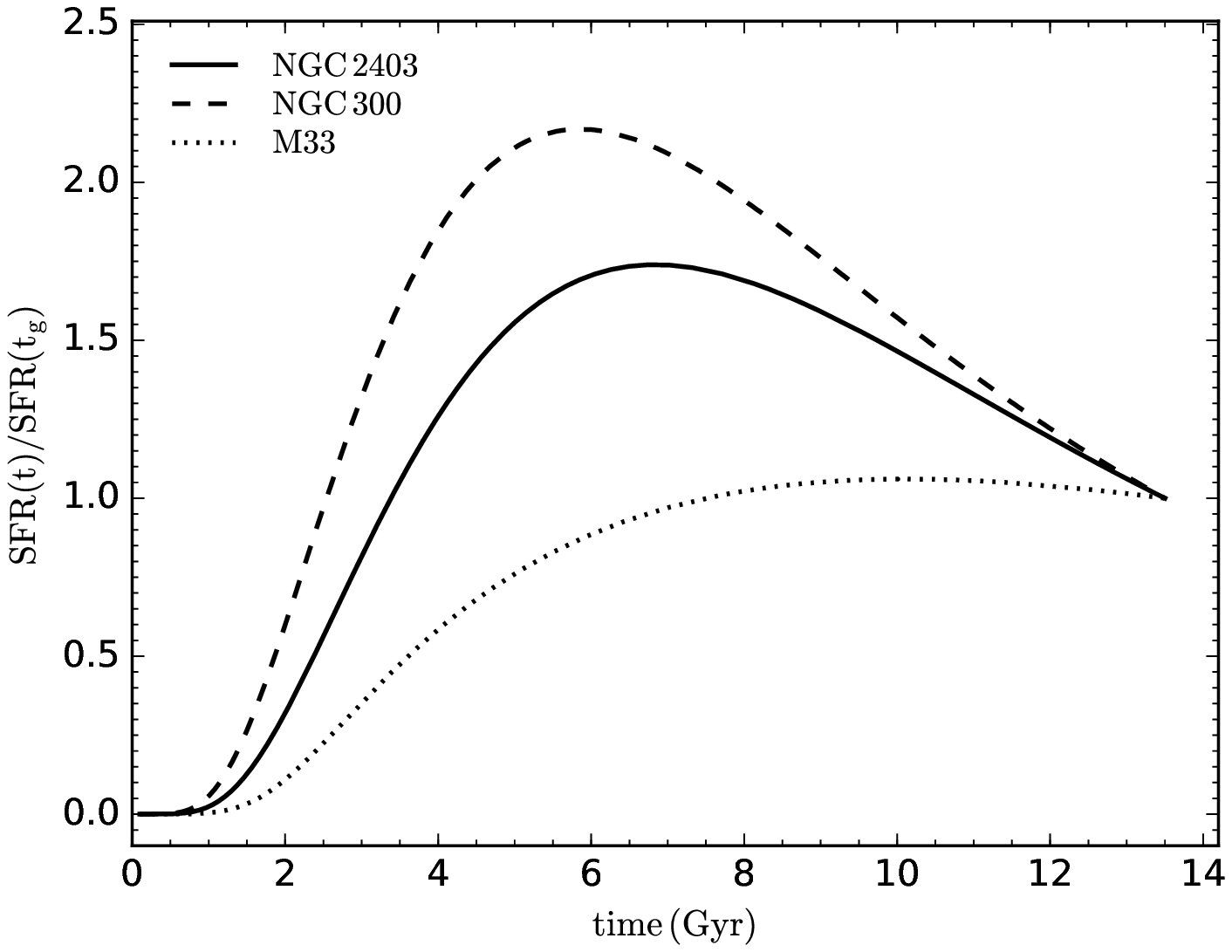}
   \includegraphics[angle=0,scale=0.55]{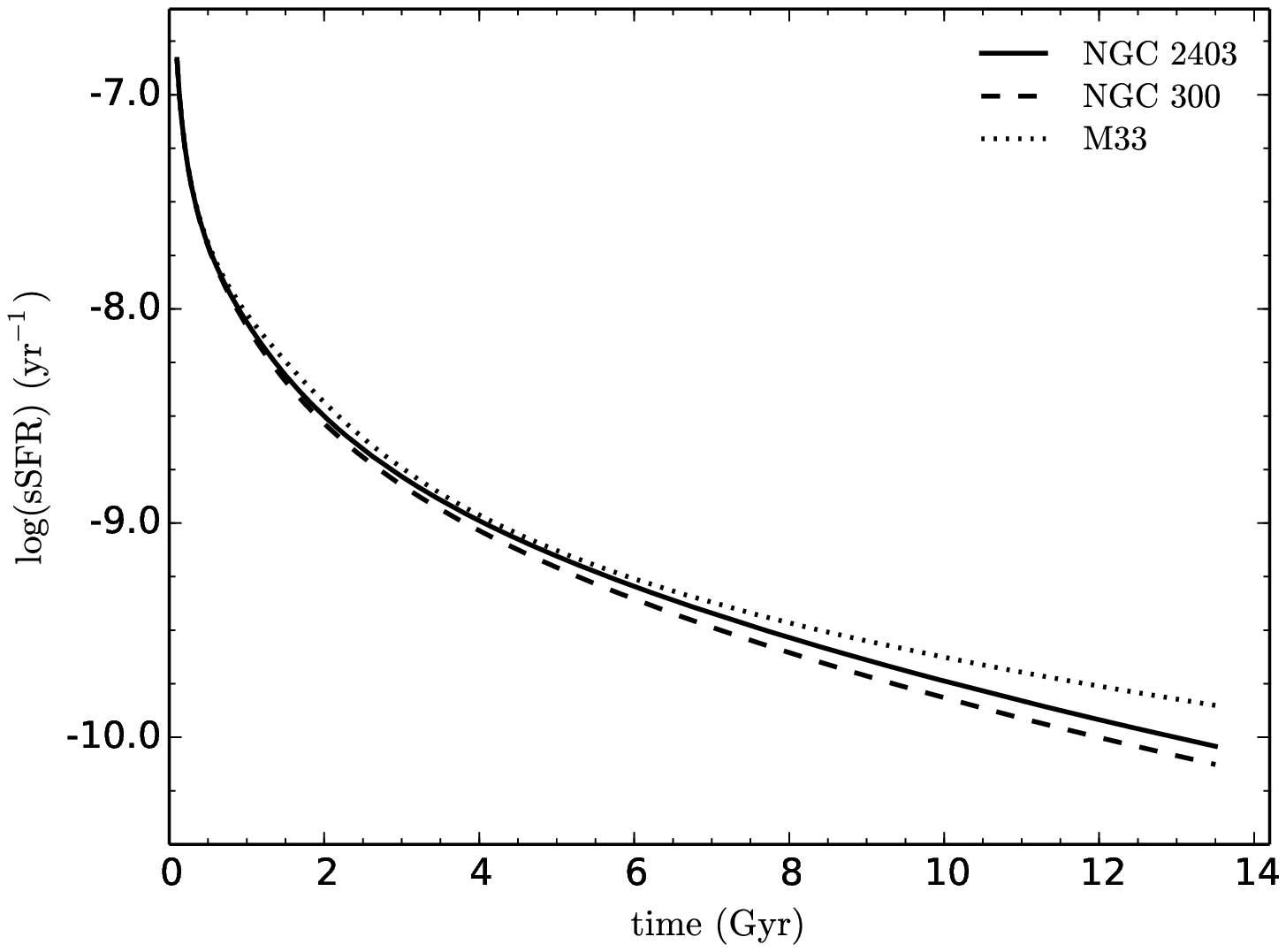}
   \caption{The evolution of SFR (left panel) and specific
   SFR (sSFR, right panel) predicted by their own best-fitting
   models. The different line types correspond to different objects:
   solid line for NGC\,2403, dashed line for NGC\,300 and dotted
   line for M33.
   }
\label{Fig:SSFR}
\end{figure*}

\begin{figure}
  \centering
  \includegraphics[angle=0,scale=0.55]{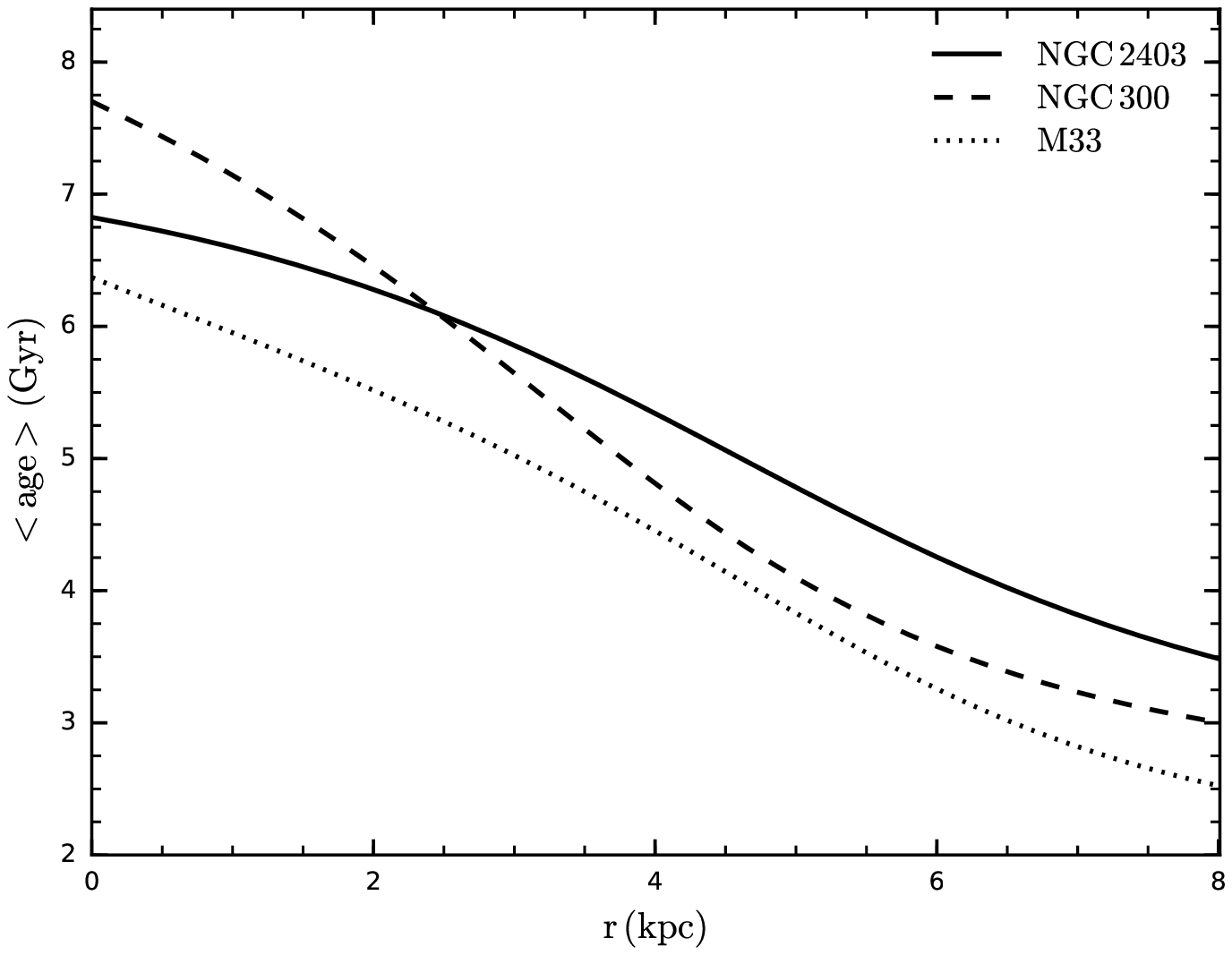}
    \caption{Comparison of the mean stellar age gradients along the discs
    of NGC\,2403 (solid line), NGC\,300 (dashed line) and M33 (dotted line)
    predicted by their own best-fitting models.
    }
  \label{Fig:age}
\end{figure}

\subsection{The evolution of the scale-length $r_{\rm d}$}

To derive the scale-length $r_{\rm d}$ of the disc at each time bin, we
fit an exponential law to the total stellar mass surface density profile.
The left panel of Fig. \ref{Fig:rd_e} shows the temporal evolution of
the disc scale-lengths $r_{\rm d}$ for the three galaxies, NGC\,2403
(solid line), NGC\,300 (dashed line) and M33 (dotted line), and the
vertical red dotted line denotes the galaxy evolutionary age at
$z\,=\,1\,({\rm i.e.,\,}t\,=\,5.75\,\rm Gyr)$. It can be seen that $r_{\rm d}$
increases with time in all cases, and the growth of the three discs
seems to be approximately linear since $z\,=\,1\,({\rm i.e.,\,}t\,=\,5.75\,\rm Gyr)$.
Then the disc scale-length growth rate $dr_{\rm d}/dt$ can be expressed as
the mean temporal disc growth rate, that is,
$dr_{\rm d}/dt\,=\,\frac{r_{\rm d}(z=0)-r_{\rm d}(z=1)}{t(z=0)-t(z=1)}$.
Therefore, the stellar disc scale-length growth rate between
$z\,=\,1$ and $z\,=\,0$ predicted by their own best-fitting models are
$0.0387\,\rm kpc\,Gyr^{-1}$ for NGC\,2403, $0.0372\,\rm kpc\,Gyr^{-1}$ for
NGC\,300 and $0.0445\,\rm kpc\,Gyr^{-1}$ for M33, fairly consistent with the
statistic results of \citet{MM11} that the scale-length of late-type discs
appear to grow at a rate of $0.02-0.04\,\rm kpc\,Gyr^{-1}$.

Rather than using the \emph{absolute} growth rate in unit of $\rm kpc\,Gyr^{-1}$
to describe the evolution of galaxy discs, it should be more illustrative
to focus on their \emph{relative} size increase. The evolution of \emph{relative}
disc scale-lengths of these three galaxies predicted by their own best fitting models are shown
in the right panel of Fig. \ref{Fig:rd_e}, normalized to their sizes at $z\,=\,1$.
It can be seen that the disc size of them show clear growth since $z\,=\,1$,
and the relative increase $r_{\rm d}(z\,=\,0)/r_{\rm d}(z\,=\,1)$ is 1.242 for NGC\,2403,
1.2778 for NGC\,300, and 1.3396 for M33, respectively. That is, compared to M33,
the scale-lengths of both NGC\,2403 and NGC\,300 have less growth \citep{Hillis2016}.
Our results are in perfect agreement with the \emph{relative} disc growth rate
estimated by \citet{MM07} and \citet{MM11}. In addition, it indicates that, even
though the absolute growth rate of NGC\,2403 is high, it will not have a
significant increase in its scale-length.

Figure \ref{Fig:MS} plots the evolution of stellar-mass--scale-length relation
from $z\,=\,1$ to $z\,=\,0$ for NGC\,2403 (solid line), NGC\,300 (dashed line)
and M33 (dotted line). The $z\,=\,0$ step and $z\,=\,1$ step is marked with
filled circle and with filled asterisk, respectively. From Figure. \ref{Fig:MS},
it can be seen that both stellar mass and scale-length simultaneously increase
as time goes by, which reinforces the result of the previous studies
\citep[e.g., ][and references therein]{brooks11, MM11, brook12}.

\subsection{The evolution of the stellar mass}

The left panel of Fig. \ref{Fig:stellar} displays the best-fitting
model predicted evolution of the stellar mass for NGC\,2403 (solid line),
NGC\,300 (dashed line)
and M33 (dotted line). To make the growth history of $M_{\ast}$ more
visible, stellar masses are normalized to their present-day values;
and to compute the $M_{\ast}$ growth rate, the horizontal dot-dashed
line in the panel marks when the stellar mass achieves 50\% of its
final value. It can be seen that all of them have
been steadily increasing to their present-day values. Furthermore,
these three galaxies gained more than 50\% of their total stellar
mass in the past $\sim\rm 8\,Gyr$ (i.e., since $z\,=\,1$), consistent
with the previous results that late-type galaxies with stellar mass
$M_{*}\,<\,10^{11}\,{\rm M}_{\odot}$ appear to gain most of their
stellar mass at $z\,<\,1$ \citep{Leitner2011, Sachdeva2015}.

To show the accumulation of stellar mass for them more vividly,
we also plot the \emph{relative} accumulated history of stellar mass
since $z\,=\,1$ (${\rm i.e.,}\,t\,=\,5.75\rm \,Gyr$) in the right
panel of Fig. \ref{Fig:stellar}, here normalized to the value at
$z\,=\,1$. The line types are the same as that in the left panel.
It can be found that all the three galaxies become more massive
with time, and we derive that the ratio between present-day
value of stellar mass and the value at $z\,=\,1$, $M_{*}(z=0)/M_{*}(z=1)$,
is $3.586$ for NGC\,2403, $2.958$ for NGC\,300, and $4.713$ for M33,
respectively. It suggests that the principal time of star formation on
the disc of M33 is later than that in NGC\,2403 and NGC\,300, indicating
that the mean age of stellar population in M33 may be younger than that
in NGC\,2403 and NGC\,300.

\subsection{The evolution of SFR and sSFR}

The SFHs of the three galaxies are displayed in the left panel of
Fig. \ref{Fig:SSFR}, and the different line types correspond to
different galaxies: solid line for NGC\,2403, dashed line for
NGC\,300 and dotted line for M33. Fig. \ref{Fig:SSFR} shows the
SFH of NGC\,2403 is similar to that of NGC\,300, and both NGC\,2403 and NGC\,300
reach their peaks earlier than M33, that is, the principal time of
star formation on the discs of NGC\,2403 and NGC\,300 is earlier than
that on the disc of M33.

The temporal evolution of sSFR is shown in right panel of Fig. \ref{Fig:SSFR},
and the line types are the same as that in the left panel. It can be found
from the right panel of Fig. \ref{Fig:SSFR}
that the sSFR of all the three galaxies decreases with time, and both
NGC\,2403 and NGC\,300 have lower sSFR than that in M33 during their
whole histories. Furthermore, the difference among them increases with time
and reaches the maximum now. All these indicate that the star formation
activity in M33 is more intense than that in other two isolated galaxies,
and both observations and simulations indeed shown that interaction
can lead to high star formation efficiencies \citep{schuster07}.
Indeed, previous authors have concluded that star formation activity in
the case of M33 is slightly different, that is, M33 is more efficient in
forming stars than other spirals in the local Universe, especially for
large spirals \citep{Gardan2007, Gratier2010}.
The corresponding values of present-day sSFR predicted by their own
best-fitting models are $-10.042\,\rm yr^{-1}$ for NGC\,2403,
$-10.127\,\rm yr^{-1}$ for NGC\,300 and $-9.841\,\rm yr^{-1}$ for M33.
Meanwhile, \citet{MM07} used FUV$-K$ color as a proxy for sSFR, and
derived that the values of sSFR for NGC\,2403, NGC\,300 and M33 are
$-10.249\,\pm\,0.025\,\rm yr^{-1}$, $-10.253\,\pm\,0.02\,\rm yr^{-1}$
and $-9.953\,\pm\,0.005\,\rm yr^{-1}$, respectively. Thus, our best-fitting
model predicted sSFR values are in good agreement with the observed
values of \citet{MM07}, considering the the observed uncertainties.

\subsection{The age gradient}

In order to display clearly the property of the stellar age
of the three galaxies, we plot the mean stellar age along the
discs of them, predicted by their own best-fitting models, with the
solid line for NGC\,2403, dashed line for NGC\,300 and dotted line
for M33 in Fig. \ref{Fig:age}. It can be seen that there exists
radial age gradient in all of them, and the age gradient of NGC\,2403
is shallower than those of NGC\,300 and M33.
The best-fitting model predicted stellar population mean age
in the central region is 6.82\,Gyr for NGC\,2403, 7.70\,Gyr for
NGC\,300 and $6.37\,\rm Gyr$ for M33, and that at $r\,=\,8.0\,\rm kpc$
is 3.44\,Gyr, 2.97\,Gyr and 2.48\,Gyr in the corresponding discs.
We obtain the age gradient of NGC\,2403 is $0.4225\,\rm Gyr\,kpc^{-1}$,
consistent with the previous observational work for the age distribution
along the disc of NGC\,2403 \citep{barker2012, williams13a}.
This further confirms the above results that the mean age of stellar
populations along the whole discs of both NGC\,2403 and NGC\,300 is
older than that in M33. Meanwhile, it also indicates that recently
there are more gas infall onto the disc of M33 and more star formation
occurred along the M33 disc than that in the other two galaxies.
This is mainly due to the fact that there exists an H{\sc i} bridge between
M33 and M31, which is responsible for the future star formation in
their discs \citep{Wolfe2013}.

\section{Summary}

In this work, we analyse the evolution and SFH for the disc of NGC\,2403 by
constructing a simple model, which is assumed that the disc of NGC\,2403
originates and
builds up through the accretion of the primordial gas, and the gas outflow
process is also considered in the model. Our results show that the model
results are very sensitive to the adopted gas infall timescale, while the
outflow process mainly determined the shape of gas-phase metallicity
distribution along the disc of NG\,2403.

To provide a picture of the local environment on the evolution of galaxies
and whether or not the isolated disc galaxies
experience similar chemical evolution history,
we compare our results of NGC\,2403 with those of nearby morphological
twins M33 and NGC\,300, which are studied in our previous
work \citep{kang12, kang16}. We found that these three galaxies gained
more than 50 percent of their stellar mass in the past $\sim\,8\,\rm Gyr$
(i.e., at $z\,<\,1$).
Our results also show that the metallicity gradient in M33 seems
to be flatter than that in isolated galaxies NGC\,2403 and NGC\,300,
and the metallicity gradient in NGC\,2403 and NGC\,300 are similar,
when the metallicity gradients are expressed in ${\rm dex}\,R_{\rm 25}^{-1}$,
which implies that both NGC\,2403 and M33 may follow similar chemical
evolutionary histories.
The principal epoch of star formation on the discs of NGC\,2403 and
NGC\,300 is earlier than that on the disc on M33, and the mean
age of stellar populations along the whole discs of both NGC\,2403
and NGC\,300 is older than that of M33. The local environment plays
an important role on the evolution and SFH of a galaxy, at least for
galaxies with stellar mass of $10^{9.2}\,\rm M_{\odot}\sim10^{9.7}\,\rm M_{\odot}$.

Although we obtain the reasonable results about the environment on
the evolution and SFH of spiral galaxies, we only focus on three
spiral galaxies (NGC\,2403, NGC\,300 and M33) in the work. The sample
size is small now, we will investigate large sample in the near future
work, and IFU surveys provide large observational sample size for our
future study.

\section*{Acknowledgements}

We thank the anonymous referee whose comments and suggestions
have improved the quality of this paper greatly.
Xiaoyu Kang and Fenghui Zhang are supported by the National
Natural Science Foundation of China (NSFC,
No. 11573062, 11403092, 11390374 and 11521303), the YIPACAS
Foundation (No. 2012048), and the Yunnan Foundation (No. 2011CI053).
Ruixiang Chang is supported by the NSFC (No. 11373053 and 11390373),
and Strategic Priority Research Program "The Emergence of
Cosmological Structures" of the Chinese Academy of Sciences
(CAS, No. XDB09010100).
Xiaoyu Kang thanks Rolf-Peter Kudritzki in
University of Hawaii for helpful suggestions during revision.




\bibliographystyle{mnras}
\bibliography{kxy} 

\begin{thebibliography}{}
\makeatletter
\relax
\def\mn@urlcharsother{\let\do\@makeother \do\$\do\&\do\#\do\^\do\_\do\%\do\~}
\def\mn@doi{\begingroup\mn@urlcharsother \@ifnextchar [ {\mn@doi@}
  {\mn@doi@[]}}
\def\mn@doi@[#1]#2{\def\@tempa{#1}\ifx\@tempa\@empty \href
  {http://dx.doi.org/#2} {doi:#2}\else \href {http://dx.doi.org/#2} {#1}\fi
  \endgroup}
\def\mn@eprint#1#2{\mn@eprint@#1:#2::\@nil}
\def\mn@eprint@arXiv#1{\href {http://arxiv.org/abs/#1} {{\tt arXiv:#1}}}
\def\mn@eprint@dblp#1{\href {http://dblp.uni-trier.de/rec/bibtex/#1.xml}
  {dblp:#1}}
\def\mn@eprint@#1:#2:#3:#4\@nil{\def\@tempa {#1}\def\@tempb {#2}\def\@tempc
  {#3}\ifx \@tempc \@empty \let \@tempc \@tempb \let \@tempb \@tempa \fi \ifx
  \@tempb \@empty \def\@tempb {arXiv}\fi \@ifundefined
  {mn@eprint@\@tempb}{\@tempb:\@tempc}{\expandafter \expandafter \csname
  mn@eprint@\@tempb\endcsname \expandafter{\@tempc}}}

\bibitem[\protect\citeauthoryear{{Asplund}, {Grevesse}, {Sauval}  \&
  {Scott}}{{Asplund} et~al.}{2009}]{asplund09}
{Asplund} M.,  {Grevesse} N.,  {Sauval} A.~J.,   {Scott} P.,  2009, \mn@doi
  [\araa] {10.1146/annurev.astro.46.060407.145222}, \href
  {http://adsabs.harvard.edu/abs/2009ARA%26A..47..481A} {47, 481}

\bibitem[\protect\citeauthoryear{{Barker}, {Sarajedini}, {Geisler}, {Harding}
  \& {Schommer}}{{Barker} et~al.}{2007}]{barker07}
{Barker} M.~K.,  {Sarajedini} A.,  {Geisler} D.,  {Harding} P.,   {Schommer}
  R.,  2007, \mn@doi [\aj] {10.1086/511185}, \href
  {http://adsabs.harvard.edu/abs/2007AJ....133.1125B} {133, 1125}

\bibitem[\protect\citeauthoryear{{Barker}, {Ferguson}, {Cole}, {Ibata},
  {Irwin}, {Lewis}, {Smecker-Hane}  \& {Tanvir}}{{Barker}
  et~al.}{2011}]{barker11}
{Barker} M.~K.,  {Ferguson} A.~M.~N.,  {Cole} A.~A.,  {Ibata} R.,  {Irwin} M.,
  {Lewis} G.~F.,  {Smecker-Hane} T.~A.,   {Tanvir} N.~R.,  2011, \mn@doi
  [\mnras] {10.1111/j.1365-2966.2010.17458.x}, \href
  {http://adsabs.harvard.edu/abs/2011MNRAS.410..504B} {410, 504}

\bibitem[\protect\citeauthoryear{{Barker}, {Ferguson}, {Irwin}, {Arimoto}  \&
  {Jablonka}}{{Barker} et~al.}{2012}]{barker2012}
{Barker} M.~K.,  {Ferguson} A.~M.~N.,  {Irwin} M.~J.,  {Arimoto} N.,
  {Jablonka} P.,  2012, \mn@doi [\mnras] {10.1111/j.1365-2966.2011.19814.x},
  \href {http://ads.bao.ac.cn/abs/2012MNRAS.419.1489B} {419, 1489}

\bibitem[\protect\citeauthoryear{{Berg}, {Skillman}, {Garnett}, {Croxall},
  {Marble}, {Smith}, {Gordon}  \& {Kennicutt}}{{Berg} et~al.}{2013}]{Berg2013}
{Berg} D.~A.,  {Skillman} E.~D.,  {Garnett} D.~R.,  {Croxall} K.~V.,  {Marble}
  A.~R.,  {Smith} J.~D.,  {Gordon} K.,   {Kennicutt} Jr. R.~C.,  2013, \mn@doi
  [\apj] {10.1088/0004-637X/775/2/128}, \href
  {http://adsabs.harvard.edu/abs/2013ApJ...775..128B} {775, 128}

\bibitem[\protect\citeauthoryear{{Bernard} et~al.,}{{Bernard}
  et~al.}{2012}]{Bernard12}
{Bernard} E.~J.,  et~al., 2012, \mn@doi [\mnras]
  {10.1111/j.1365-2966.2011.20234.x}, \href
  {http://adsabs.harvard.edu/abs/2012MNRAS.420.2625B} {420, 2625}

\bibitem[\protect\citeauthoryear{{Bigiel}, {Cormier}  \& {Schmidt}}{{Bigiel}
  et~al.}{2014}]{Bigiel2014}
{Bigiel} F.,  {Cormier} D.,   {Schmidt} T.,  2014, \mn@doi [Astronomische
  Nachrichten] {10.1002/asna.201412067}, \href
  {http://adsabs.harvard.edu/abs/2014AN....335..470B} {335, 470}

\bibitem[\protect\citeauthoryear{{Bland-Hawthorn}, {Vlaji{\'c}}, {Freeman}  \&
  {Draine}}{{Bland-Hawthorn} et~al.}{2005}]{bland05}
{Bland-Hawthorn} J.,  {Vlaji{\'c}} M.,  {Freeman} K.~C.,   {Draine} B.~T.,
  2005, \mn@doi [\apj] {10.1086/430512}, \href
  {http://adsabs.harvard.edu/abs/2005ApJ...629..239B} {629, 239}

\bibitem[\protect\citeauthoryear{{Boissier} \& {Prantzos}}{{Boissier} \&
  {Prantzos}}{2000}]{bp00}
{Boissier} S.,  {Prantzos} N.,  2000, \mn@doi [\mnras]
  {10.1046/j.1365-8711.2000.03133.x}, \href
  {http://adsabs.harvard.edu/abs/2000MNRAS.312..398B} {312, 398}

\bibitem[\protect\citeauthoryear{{Braun} \& {Thilker}}{{Braun} \&
  {Thilker}}{2004}]{braun04}
{Braun} R.,  {Thilker} D.~A.,  2004, \mn@doi [\aap]
  {10.1051/0004-6361:20034423}, \href
  {http://adsabs.harvard.edu/abs/2004A%26A...417..421B} {417, 421}

\bibitem[\protect\citeauthoryear{{Bresolin}, {Kennicutt}  \&
  {Garnett}}{{Bresolin} et~al.}{1999}]{Bresolin1999}
{Bresolin} F.,  {Kennicutt} Jr. R.~C.,   {Garnett} D.~R.,  1999, \mn@doi [\apj]
  {10.1086/306576}, \href {http://adsabs.harvard.edu/abs/1999ApJ...510..104B}
  {510, 104}

\bibitem[\protect\citeauthoryear{{Bresolin}, {Gieren}, {Kudritzki},
  {Pietrzy{\'n}ski}, {Urbaneja}  \& {Carraro}}{{Bresolin}
  et~al.}{2009}]{bresolin09}
{Bresolin} F.,  {Gieren} W.,  {Kudritzki} R.-P.,  {Pietrzy{\'n}ski} G.,
  {Urbaneja} M.~A.,   {Carraro} G.,  2009, \mn@doi [\apj]
  {10.1088/0004-637X/700/1/309}, \href
  {http://adsabs.harvard.edu/abs/2009ApJ...700..309B} {700, 309}

\bibitem[\protect\citeauthoryear{{Brook} et~al.,}{{Brook}
  et~al.}{2012}]{brook12}
{Brook} C.~B.,  et~al., 2012, \mn@doi [\mnras]
  {10.1111/j.1365-2966.2012.21738.x}, \href
  {http://adsabs.harvard.edu/abs/2012MNRAS.426..690B} {426, 690}

\bibitem[\protect\citeauthoryear{{Brooks} et~al.,}{{Brooks}
  et~al.}{2011}]{brooks11}
{Brooks} A.~M.,  et~al., 2011, \mn@doi [\apj] {10.1088/0004-637X/728/1/51},
  \href {http://adsabs.harvard.edu/abs/2011ApJ...728...51B} {728, 51}

\bibitem[\protect\citeauthoryear{{Byun} \& {Freeman}}{{Byun} \&
  {Freeman}}{1995}]{B&F95}
{Byun} Y.~I.,  {Freeman} K.~C.,  1995, \mn@doi [\apj] {10.1086/175986}, \href
  {http://adsabs.harvard.edu/abs/1995ApJ...448..563B} {448, 563}

\bibitem[\protect\citeauthoryear{{Cepa}, {Prieto}, {Beckman}  \&
  {Munoz-Tunon}}{{Cepa} et~al.}{1988}]{Cepa1988}
{Cepa} J.,  {Prieto} M.,  {Beckman} J.,   {Munoz-Tunon} C.,  1988, \aap, \href
  {http://ads.bao.ac.cn/abs/1988A%26A...193...15C} {193, 15}

\bibitem[\protect\citeauthoryear{{Chabrier}}{{Chabrier}}{2003}]{chabrier03}
{Chabrier} G.,  2003, \mn@doi [\apjl] {10.1086/374879}, \href
  {http://adsabs.harvard.edu/abs/2003ApJ...586L.133C} {586, L133}

\bibitem[\protect\citeauthoryear{{Chang}, {Hou}, {Shu}  \& {Fu}}{{Chang}
  et~al.}{1999}]{chang99}
{Chang} R.~X.,  {Hou} J.~L.,  {Shu} C.~G.,   {Fu} C.~Q.,  1999, \aap, \href
  {http://adsabs.harvard.edu/abs/1999A%26A...350...38C} {350, 38}

\bibitem[\protect\citeauthoryear{{Chang}, {Hou}, {Shen}  \& {Shu}}{{Chang}
  et~al.}{2010}]{chang10}
{Chang} R.~X.,  {Hou} J.~L.,  {Shen} S.~Y.,   {Shu} C.~G.,  2010, \mn@doi
  [\apj] {10.1088/0004-637X/722/1/380}, \href
  {http://adsabs.harvard.edu/abs/2010ApJ...722..380C} {722, 380}

\bibitem[\protect\citeauthoryear{{Chang}, {Shen}  \& {Hou}}{{Chang}
  et~al.}{2012}]{chang12}
{Chang} R.~X.,  {Shen} S.~Y.,   {Hou} J.~L.,  2012, \mn@doi [\apjl]
  {10.1088/2041-8205/753/1/L10}, \href
  {http://adsabs.harvard.edu/abs/2012ApJ...753L..10C} {753, L10}

\bibitem[\protect\citeauthoryear{{Chiappini}, {Matteucci}  \&
  {Romano}}{{Chiappini} et~al.}{2001}]{chiappini01}
{Chiappini} C.,  {Matteucci} F.,   {Romano} D.,  2001, \mn@doi [\apj]
  {10.1086/321427}, \href {http://adsabs.harvard.edu/abs/2001ApJ...554.1044C}
  {554, 1044}

\bibitem[\protect\citeauthoryear{{Chiosi}}{{Chiosi}}{1980}]{chiosi1980}
{Chiosi} C.,  1980, \aap, \href
  {http://adsabs.harvard.edu/abs/1980A%26A....83..206C} {83, 206}

\bibitem[\protect\citeauthoryear{{Cooper} et~al.,}{{Cooper}
  et~al.}{2006}]{Cooper2006}
{Cooper} M.~C.,  et~al., 2006, \mn@doi [\mnras]
  {10.1111/j.1365-2966.2006.10485.x}, \href
  {http://adsabs.harvard.edu/abs/2006MNRAS.370..198C} {370, 198}

\bibitem[\protect\citeauthoryear{{Cooper}, {Tremonti}, {Newman}  \&
  {Zabludoff}}{{Cooper} et~al.}{2008}]{Cooper2008}
{Cooper} M.~C.,  {Tremonti} C.~A.,  {Newman} J.~A.,   {Zabludoff} A.~I.,  2008,
  \mn@doi [\mnras] {10.1111/j.1365-2966.2008.13714.x}, \href
  {http://adsabs.harvard.edu/abs/2008MNRAS.390..245C} {390, 245}

\bibitem[\protect\citeauthoryear{{Dalcanton}, {Williams}, {Seth}, {Dolphin},
  {Holtzman}, {Rosema}, {Skillman}  \& {Quinn}}{{Dalcanton}
  et~al.}{2009}]{Dalcanton09}
{Dalcanton} J.~J.,  {Williams} B.~F.,  {Seth} A.~C.,  {Dolphin} A.,  {Holtzman}
  J.,  {Rosema} K.,  {Skillman} E.~D.,   {Quinn} T.,  2009, \mn@doi [\apjs]
  {10.1088/0067-0049/183/1/67}, \href
  {http://adsabs.harvard.edu/abs/2009ApJS..183...67D} {183, 67}

\bibitem[\protect\citeauthoryear{{Davis} \& {Geller}}{{Davis} \&
  {Geller}}{1976}]{Davis1976}
{Davis} M.,  {Geller} M.~J.,  1976, \mn@doi [\apj] {10.1086/154575}, \href
  {http://adsabs.harvard.edu/abs/1976ApJ...208...13D} {208, 13}

\bibitem[\protect\citeauthoryear{{Dressler}}{{Dressler}}{1980}]{Dressler1980}
{Dressler} A.,  1980, \mn@doi [\apj] {10.1086/157753}, \href
  {http://adsabs.harvard.edu/abs/1980ApJ...236..351D} {236, 351}

\bibitem[\protect\citeauthoryear{{Ellison}, {Simard}, {Cowan}, {Baldry},
  {Patton}  \& {McConnachie}}{{Ellison} et~al.}{2009}]{Ellison2009}
{Ellison} S.~L.,  {Simard} L.,  {Cowan} N.~B.,  {Baldry} I.~K.,  {Patton}
  D.~R.,   {McConnachie} A.~W.,  2009, \mn@doi [\mnras]
  {10.1111/j.1365-2966.2009.14817.x}, \href
  {http://adsabs.harvard.edu/abs/2009MNRAS.396.1257E} {396, 1257}

\bibitem[\protect\citeauthoryear{{Ferguson}, {Irwin2}, {Chapman}, {Ibata},
  {Lewis}  \& {Tanvir}}{{Ferguson} et~al.}{2007}]{Ferguson07}
{Ferguson} A.,  {Irwin2} M.,  {Chapman} S.,  {Ibata} R.,  {Lewis} G.,
  {Tanvir} N.,  2007, {Resolving the Stellar Outskirts of M31 and M33}.
p.~239, \mn@doi{10.1007/978-1-4020-5573-7_39}

\bibitem[\protect\citeauthoryear{{Freedman} et~al.,}{{Freedman}
  et~al.}{2001}]{Freedman2001}
{Freedman} W.~L.,  et~al., 2001, \mn@doi [\apj] {10.1086/320638}, \href
  {http://adsabs.harvard.edu/abs/2001ApJ...553...47F} {553, 47}

\bibitem[\protect\citeauthoryear{{Gardan}, {Braine}, {Schuster}, {Brouillet}
  \& {Sievers}}{{Gardan} et~al.}{2007}]{Gardan2007}
{Gardan} E.,  {Braine} J.,  {Schuster} K.~F.,  {Brouillet} N.,   {Sievers} A.,
  2007, \mn@doi [\aap] {10.1051/0004-6361:20077711}, \href
  {http://adsabs.harvard.edu/abs/2007A%26A...473...91G} {473, 91}

\bibitem[\protect\citeauthoryear{{Garnett}}{{Garnett}}{2002}]{garnett02}
{Garnett} D.~R.,  2002, \mn@doi [\apj] {10.1086/344301}, \href
  {http://adsabs.harvard.edu/abs/2002ApJ...581.1019G} {581, 1019}

\bibitem[\protect\citeauthoryear{{Garnett}, {Shields}, {Skillman}, {Sagan}  \&
  {Dufour}}{{Garnett} et~al.}{1997}]{Garnett1997}
{Garnett} D.~R.,  {Shields} G.~A.,  {Skillman} E.~D.,  {Sagan} S.~P.,
  {Dufour} R.~J.,  1997, \apj, \href
  {http://adsabs.harvard.edu/abs/1997ApJ...489...63G} {489, 63}

\bibitem[\protect\citeauthoryear{{Garnett}, {Shields}, {Peimbert},
  {Torres-Peimbert}, {Skillman}, {Dufour}, {Terlevich}  \&
  {Terlevich}}{{Garnett} et~al.}{1999}]{Garnett1999}
{Garnett} D.~R.,  {Shields} G.~A.,  {Peimbert} M.,  {Torres-Peimbert} S.,
  {Skillman} E.~D.,  {Dufour} R.~J.,  {Terlevich} E.,   {Terlevich} R.~J.,
  1999, \mn@doi [\apj] {10.1086/306860}, \href
  {http://adsabs.harvard.edu/abs/1999ApJ...513..168G} {513, 168}

\bibitem[\protect\citeauthoryear{{Goddard} et~al.,}{{Goddard}
  et~al.}{2017}]{Goddard2016}
{Goddard} D.,  et~al., 2017, \mn@doi [\mnras] {10.1093/mnras/stw2719}, \href
  {http://adsabs.harvard.edu/abs/2017MNRAS.465..688G} {465, 688}

\bibitem[\protect\citeauthoryear{{Gogarten} et~al.,}{{Gogarten}
  et~al.}{2010}]{gogarten10}
{Gogarten} S.~M.,  et~al., 2010, \mn@doi [\apj] {10.1088/0004-637X/712/2/858},
  \href {http://adsabs.harvard.edu/abs/2010ApJ...712..858G} {712, 858}

\bibitem[\protect\citeauthoryear{{Gratier} et~al.,}{{Gratier}
  et~al.}{2010}]{Gratier2010}
{Gratier} P.,  et~al., 2010, \mn@doi [\aap] {10.1051/0004-6361/201014441},
  \href {http://adsabs.harvard.edu/abs/2010A%26A...522A...3G} {522, A3}

\bibitem[\protect\citeauthoryear{{Gupta}, {Yuan}, {Tran}, {Martizzi}, {Taylor}
  \& {Kewley}}{{Gupta} et~al.}{2016}]{Gupta2016}
{Gupta} A.,  {Yuan} T.,  {Tran} K.-V.~H.,  {Martizzi} D.,  {Taylor} P.,
  {Kewley} L.~J.,  2016, \mn@doi [\apj] {10.3847/0004-637X/831/1/104}, \href
  {http://adsabs.harvard.edu/abs/2016ApJ...831..104G} {831, 104}

\bibitem[\protect\citeauthoryear{{Heesen}, {Brinks}, {Leroy}, {Heald}, {Braun},
  {Bigiel}  \& {Beck}}{{Heesen} et~al.}{2014}]{heesen14}
{Heesen} V.,  {Brinks} E.,  {Leroy} A.~K.,  {Heald} G.,  {Braun} R.,  {Bigiel}
  F.,   {Beck} R.,  2014, \mn@doi [\aj] {10.1088/0004-6256/147/5/103}, \href
  {http://adsabs.harvard.edu/abs/2014AJ....147..103H} {147, 103}

\bibitem[\protect\citeauthoryear{{Henry} \& {Worthey}}{{Henry} \&
  {Worthey}}{1999}]{Henry1999}
{Henry} R.~B.~C.,  {Worthey} G.,  1999, \mn@doi [\pasp] {10.1086/316403}, \href
  {http://adsabs.harvard.edu/abs/1999PASP..111..919H} {111, 919}

\bibitem[\protect\citeauthoryear{{Hillis}, {Williams}, {Dolphin}, {Dalcanton}
  \& {Skillman}}{{Hillis} et~al.}{2016}]{Hillis2016}
{Hillis} T.~J.,  {Williams} B.~F.,  {Dolphin} A.~E.,  {Dalcanton} J.~J.,
  {Skillman} E.~D.,  2016, \mn@doi [\apj] {10.3847/0004-637X/831/2/191}, \href
  {http://adsabs.harvard.edu/abs/2016ApJ...831..191H} {831, 191}

\bibitem[\protect\citeauthoryear{{Ho}, {Kudritzki}, {Kewley}, {Zahid},
  {Dopita}, {Bresolin}  \& {Rupke}}{{Ho} et~al.}{2015}]{ho15}
{Ho} I.-T.,  {Kudritzki} R.-P.,  {Kewley} L.~J.,  {Zahid} H.~J.,  {Dopita}
  M.~A.,  {Bresolin} F.,   {Rupke} D.~S.~N.,  2015, \mn@doi [\mnras]
  {10.1093/mnras/stv067}, \href
  {http://adsabs.harvard.edu/abs/2015MNRAS.448.2030H} {448, 2030}

\bibitem[\protect\citeauthoryear{{Hughes}, {Cortese}, {Boselli}, {Gavazzi}  \&
  {Davies}}{{Hughes} et~al.}{2013}]{hughes13}
{Hughes} T.~M.,  {Cortese} L.,  {Boselli} A.,  {Gavazzi} G.,   {Davies} J.~I.,
  2013, \mn@doi [\aap] {10.1051/0004-6361/201218822}, \href
  {http://adsabs.harvard.edu/abs/2013A%26A...550A.115H} {550, A115}

\bibitem[\protect\citeauthoryear{{Hunter} \& {Thronson}}{{Hunter} \&
  {Thronson}}{1995}]{H&T1995}
{Hunter} D.~A.,  {Thronson} Jr. H.~A.,  1995, \mn@doi [\apj] {10.1086/176295},
  \href {http://adsabs.harvard.edu/abs/1995ApJ...452..238H} {452, 238}

\bibitem[\protect\citeauthoryear{{Iovino} et~al.,}{{Iovino}
  et~al.}{2010}]{Iovino2010}
{Iovino} A.,  et~al., 2010, \mn@doi [\aap] {10.1051/0004-6361/200912558}, \href
  {http://adsabs.harvard.edu/abs/2010A%26A...509A..40I} {509, A40}

\bibitem[\protect\citeauthoryear{{Jarrett}, {Chester}, {Cutri}, {Schneider}  \&
  {Huchra}}{{Jarrett} et~al.}{2003}]{jarrett03}
{Jarrett} T.~H.,  {Chester} T.,  {Cutri} R.,  {Schneider} S.~E.,   {Huchra}
  J.~P.,  2003, \mn@doi [\aj] {10.1086/345794}, \href
  {http://adsabs.harvard.edu/abs/2003AJ....125..525J} {125, 525}

\bibitem[\protect\citeauthoryear{{Kang}, {Chang}, {Yin}, {Hou}, {Zhang},
  {Zhang}  \& {Han}}{{Kang} et~al.}{2012}]{kang12}
{Kang} X.,  {Chang} R.,  {Yin} J.,  {Hou} J.,  {Zhang} F.,  {Zhang} Y.,   {Han}
  Z.,  2012, \mn@doi [\mnras] {10.1111/j.1365-2966.2012.21778.x}, \href
  {http://adsabs.harvard.edu/abs/2012MNRAS.426.1455K} {426, 1455}

\bibitem[\protect\citeauthoryear{{Kang}, {Zhang}, {Chang}, {Wang}  \&
  {Cheng}}{{Kang} et~al.}{2016}]{kang16}
{Kang} X.,  {Zhang} F.,  {Chang} R.,  {Wang} L.,   {Cheng} L.,  2016, \mn@doi
  [\aap] {10.1051/0004-6361/201527041}, \href
  {http://adsabs.harvard.edu/abs/2016A%26A...585A..20K} {585, A20}

\bibitem[\protect\citeauthoryear{{Karachentsev} \& {Kaisina}}{{Karachentsev} \&
  {Kaisina}}{2013}]{K&K13}
{Karachentsev} I.~D.,  {Kaisina} E.~I.,  2013, \mn@doi [\aj]
  {10.1088/0004-6256/146/3/46}, \href
  {http://adsabs.harvard.edu/abs/2013AJ....146...46K} {146, 46}

\bibitem[\protect\citeauthoryear{{Karachentsev}, {Karachentseva}, {Huchtmeier}
  \& {Makarov}}{{Karachentsev} et~al.}{2004}]{karachentsev04}
{Karachentsev} I.~D.,  {Karachentseva} V.~E.,  {Huchtmeier} W.~K.,   {Makarov}
  D.~I.,  2004, \mn@doi [\aj] {10.1086/382905}, \href
  {http://adsabs.harvard.edu/abs/2004AJ....127.2031K} {127, 2031}

\bibitem[\protect\citeauthoryear{{Kennicutt}}{{Kennicutt}}{1983}]{Kennicutt198%
3}
{Kennicutt} Jr. R.~C.,  1983, \mn@doi [\aj] {10.1086/113334}, \href
  {http://adsabs.harvard.edu/abs/1983AJ.....88..483K} {88, 483}

\bibitem[\protect\citeauthoryear{{Kennicutt} Jr. et~al.,}{{Kennicutt}
  et~al.}{2003}]{kennicutt03}
{Kennicutt} Jr. R.~C.,  et~al., 2003, \mn@doi [\pasp] {10.1086/376941}, \href
  {http://adsabs.harvard.edu/abs/2003PASP..115..928K} {115, 928}

\bibitem[\protect\citeauthoryear{{Korotin}, {Andrievsky}, {Luck}, {L{\'e}pine},
  {Maciel}  \& {Kovtyukh}}{{Korotin} et~al.}{2014}]{korotin14}
{Korotin} S.~A.,  {Andrievsky} S.~M.,  {Luck} R.~E.,  {L{\'e}pine} J.~R.~D.,
  {Maciel} W.~J.,   {Kovtyukh} V.~V.,  2014, \mn@doi [\mnras]
  {10.1093/mnras/stu1643}, \href
  {http://adsabs.harvard.edu/abs/2014MNRAS.444.3301K} {444, 3301}

\bibitem[\protect\citeauthoryear{{Kubryk}, {Prantzos}  \&
  {Athanassoula}}{{Kubryk} et~al.}{2015}]{Kubryk2015}
{Kubryk} M.,  {Prantzos} N.,   {Athanassoula} E.,  2015, \mn@doi [\aap]
  {10.1051/0004-6361/201424171}, \href
  {http://adsabs.harvard.edu/abs/2015A%26A...580A.126K} {580, A126}

\bibitem[\protect\citeauthoryear{{Kudritzki}, {Ho}, {Schruba}, {Burkert},
  {Zahid}, {Bresolin}  \& {Dima}}{{Kudritzki} et~al.}{2015}]{Kudritzki2015}
{Kudritzki} R.-P.,  {Ho} I.-T.,  {Schruba} A.,  {Burkert} A.,  {Zahid} H.~J.,
  {Bresolin} F.,   {Dima} G.~I.,  2015, \mn@doi [\mnras]
  {10.1093/mnras/stv522}, \href
  {http://adsabs.harvard.edu/abs/2015MNRAS.450..342K} {450, 342}

\bibitem[\protect\citeauthoryear{{Lee} et~al.,}{{Lee} et~al.}{2011}]{Lee2011}
{Lee} J.~C.,  et~al., 2011, \mn@doi [\apjs] {10.1088/0067-0049/192/1/6}, \href
  {http://adsabs.harvard.edu/abs/2011ApJS..192....6L} {192, 6}

\bibitem[\protect\citeauthoryear{{Leitner} \& {Kravtsov}}{{Leitner} \&
  {Kravtsov}}{2011}]{Leitner2011}
{Leitner} S.~N.,  {Kravtsov} A.~V.,  2011, \mn@doi [\apj]
  {10.1088/0004-637X/734/1/48}, \href
  {http://adsabs.harvard.edu/abs/2011ApJ...734...48L} {734, 48}

\bibitem[\protect\citeauthoryear{{Leroy}, {Walter}, {Brinks}, {Bigiel}, {de
  Blok}, {Madore}  \& {Thornley}}{{Leroy} et~al.}{2008}]{leroy08}
{Leroy} A.~K.,  {Walter} F.,  {Brinks} E.,  {Bigiel} F.,  {de Blok} W.~J.~G.,
  {Madore} B.,   {Thornley} M.~D.,  2008, \mn@doi [\aj]
  {10.1088/0004-6256/136/6/2782}, \href
  {http://adsabs.harvard.edu/abs/2008AJ....136.2782L} {136, 2782}

\bibitem[\protect\citeauthoryear{{Leroy} et~al.,}{{Leroy}
  et~al.}{2013}]{leroy13}
{Leroy} A.~K.,  et~al., 2013, \mn@doi [\aj] {10.1088/0004-6256/146/2/19}, \href
  {http://adsabs.harvard.edu/abs/2013AJ....146...19L} {146, 19}

\bibitem[\protect\citeauthoryear{{Matteucci} \& {Francois}}{{Matteucci} \&
  {Francois}}{1989}]{matteucci89}
{Matteucci} F.,  {Francois} P.,  1989, \mnras, \href
  {http://adsabs.harvard.edu/abs/1989MNRAS.239..885M} {239, 885}

\bibitem[\protect\citeauthoryear{{McCall}, {Rybski}  \& {Shields}}{{McCall}
  et~al.}{1985}]{McCall1985}
{McCall} M.~L.,  {Rybski} P.~M.,   {Shields} G.~A.,  1985, \mn@doi [\apjs]
  {10.1086/190994}, \href {http://adsabs.harvard.edu/abs/1985ApJS...57....1M}
  {57, 1}

\bibitem[\protect\citeauthoryear{{Moll{\'a}} \& {D{\'{\i}}az}}{{Moll{\'a}} \&
  {D{\'{\i}}az}}{2005}]{m&d05}
{Moll{\'a}} M.,  {D{\'{\i}}az} A.~I.,  2005, \mn@doi [\mnras]
  {10.1111/j.1365-2966.2005.08782.x}, \href
  {http://adsabs.harvard.edu/abs/2005MNRAS.358..521M} {358, 521}

\bibitem[\protect\citeauthoryear{{Mouhcine}, {Baldry}  \& {Bamford}}{{Mouhcine}
  et~al.}{2007}]{Mouhcine2007}
{Mouhcine} M.,  {Baldry} I.~K.,   {Bamford} S.~P.,  2007, \mn@doi [\mnras]
  {10.1111/j.1365-2966.2007.12405.x}, \href
  {http://adsabs.harvard.edu/abs/2007MNRAS.382..801M} {382, 801}

\bibitem[\protect\citeauthoryear{{Moustakas}, {Kennicutt}, {Tremonti}, {Dale},
  {Smith}  \& {Calzetti}}{{Moustakas} et~al.}{2010}]{moustakas10}
{Moustakas} J.,  {Kennicutt} Jr. R.~C.,  {Tremonti} C.~A.,  {Dale} D.~A.,
  {Smith} J.-D.~T.,   {Calzetti} D.,  2010, \mn@doi [\apjs]
  {10.1088/0067-0049/190/2/233}, \href
  {http://adsabs.harvard.edu/abs/2010ApJS..190..233M} {190, 233}

\bibitem[\protect\citeauthoryear{{Mu{\~n}oz-Mateos}, {Gil de Paz}, {Boissier},
  {Zamorano}, {Jarrett}, {Gallego}  \& {Madore}}{{Mu{\~n}oz-Mateos}
  et~al.}{2007}]{MM07}
{Mu{\~n}oz-Mateos} J.~C.,  {Gil de Paz} A.,  {Boissier} S.,  {Zamorano} J.,
  {Jarrett} T.,  {Gallego} J.,   {Madore} B.~F.,  2007, \mn@doi [\apj]
  {10.1086/511812}, \href {http://adsabs.harvard.edu/abs/2007ApJ...658.1006M}
  {658, 1006}

\bibitem[\protect\citeauthoryear{{Mu{\~n}oz-Mateos}, {Boissier}, {Gil de Paz},
  {Zamorano}, {Kennicutt}, {Moustakas}, {Prantzos}  \&
  {Gallego}}{{Mu{\~n}oz-Mateos} et~al.}{2011}]{MM11}
{Mu{\~n}oz-Mateos} J.~C.,  {Boissier} S.,  {Gil de Paz} A.,  {Zamorano} J.,
  {Kennicutt} Jr. R.~C.,  {Moustakas} J.,  {Prantzos} N.,   {Gallego} J.,
  2011, \mn@doi [\apj] {10.1088/0004-637X/731/1/10}, \href
  {http://adsabs.harvard.edu/abs/2011ApJ...731...10M} {731, 10}

\bibitem[\protect\citeauthoryear{{Pilyugin} \& {Thuan}}{{Pilyugin} \&
  {Thuan}}{2005}]{PT05}
{Pilyugin} L.~S.,  {Thuan} T.~X.,  2005, \mn@doi [\apj] {10.1086/432408}, \href
  {http://adsabs.harvard.edu/abs/2005ApJ...631..231P} {631, 231}

\bibitem[\protect\citeauthoryear{{Pilyugin}, {Grebel}  \& {Kniazev}}{{Pilyugin}
  et~al.}{2014}]{pilyugin14_1}
{Pilyugin} L.~S.,  {Grebel} E.~K.,   {Kniazev} A.~Y.,  2014, \mn@doi [\aj]
  {10.1088/0004-6256/147/6/131}, \href
  {http://adsabs.harvard.edu/abs/2014AJ....147..131P} {147, 131}

\bibitem[\protect\citeauthoryear{{Pilyugin}, {Grebel}, {Zinchenko}, {Nefedyev}
  \& {Mattsson}}{{Pilyugin} et~al.}{2017}]{Pilyugin2016}
{Pilyugin} L.~S.,  {Grebel} E.~K.,  {Zinchenko} I.~A.,  {Nefedyev} Y.~A.,
  {Mattsson} L.,  2017, \mn@doi [\mnras] {10.1093/mnras/stw2831}, \href
  {http://adsabs.harvard.edu/abs/2017MNRAS.465.1358P} {465, 1358}

\bibitem[\protect\citeauthoryear{{Poggianti} et~al.,}{{Poggianti}
  et~al.}{2006}]{Poggianti2006}
{Poggianti} B.~M.,  et~al., 2006, \mn@doi [\apj] {10.1086/500666}, \href
  {http://adsabs.harvard.edu/abs/2006ApJ...642..188P} {642, 188}

\bibitem[\protect\citeauthoryear{{Prantzos} \& {Boissier}}{{Prantzos} \&
  {Boissier}}{2000}]{PB2000}
{Prantzos} N.,  {Boissier} S.,  2000, \mn@doi [\mnras]
  {10.1046/j.1365-8711.2000.03228.x}, \href
  {http://adsabs.harvard.edu/abs/2000MNRAS.313..338P} {313, 338}

\bibitem[\protect\citeauthoryear{{Puche}, {Carignan}  \& {Bosma}}{{Puche}
  et~al.}{1990}]{puche90}
{Puche} D.,  {Carignan} C.,   {Bosma} A.,  1990, \mn@doi [\aj]
  {10.1086/115612}, \href {http://adsabs.harvard.edu/abs/1990AJ....100.1468P}
  {100, 1468}

\bibitem[\protect\citeauthoryear{{Putman} et~al.,}{{Putman}
  et~al.}{2009}]{putman09}
{Putman} M.~E.,  et~al., 2009, \mn@doi [\apj] {10.1088/0004-637X/703/2/1486},
  \href {http://adsabs.harvard.edu/abs/2009ApJ...703.1486P} {703, 1486}

\bibitem[\protect\citeauthoryear{{Recchi}, {Spitoni}, {Matteucci}  \&
  {Lanfranchi}}{{Recchi} et~al.}{2008}]{recchi08}
{Recchi} S.,  {Spitoni} E.,  {Matteucci} F.,   {Lanfranchi} G.~A.,  2008,
  \mn@doi [\aap] {10.1051/0004-6361:200809879}, \href
  {http://adsabs.harvard.edu/abs/2008A%26A...489..555R} {489, 555}

\bibitem[\protect\citeauthoryear{{Rosolowsky} \& {Simon}}{{Rosolowsky} \&
  {Simon}}{2008}]{RS08}
{Rosolowsky} E.,  {Simon} J.~D.,  2008, \mn@doi [\apj] {10.1086/527407}, \href
  {http://adsabs.harvard.edu/abs/2008ApJ...675.1213R} {675, 1213}

\bibitem[\protect\citeauthoryear{{Rupke}, {Kewley}  \& {Chien}}{{Rupke}
  et~al.}{2010}]{Rupke2010}
{Rupke} D.~S.~N.,  {Kewley} L.~J.,   {Chien} L.-H.,  2010, \mn@doi [\apj]
  {10.1088/0004-637X/723/2/1255}, \href
  {http://adsabs.harvard.edu/abs/2010ApJ...723.1255R} {723, 1255}

\bibitem[\protect\citeauthoryear{{Sachdeva}, {Gadotti}, {Saha}  \&
  {Singh}}{{Sachdeva} et~al.}{2015}]{Sachdeva2015}
{Sachdeva} S.,  {Gadotti} D.~A.,  {Saha} K.,   {Singh} H.~P.,  2015, \mn@doi
  [\mnras] {10.1093/mnras/stv931}, \href
  {http://adsabs.harvard.edu/abs/2015MNRAS.451....2S} {451, 2}

\bibitem[\protect\citeauthoryear{{S{\'a}nchez}, {Rosales-Ortega}, {Jungwiert},
  {Iglesias-P{\'a}ramo}, {V{\'{\i}}lchez}, {Marino}, {Bomans}  \& {Califa
  Collaboration}}{{S{\'a}nchez} et~al.}{2013}]{sanchez13}
{S{\'a}nchez} S.~F.,  {Rosales-Ortega} F.~F.,  {Jungwiert} B.,
  {Iglesias-P{\'a}ramo} J.,  {V{\'{\i}}lchez} J.~M.,  {Marino} R.~A.,  {Bomans}
  D.,   {Califa Collaboration} 2013, \mn@doi [\aap]
  {10.1051/0004-6361/201220669}, \href
  {http://adsabs.harvard.edu/abs/2013A%26A...554A..58S} {554, A58}

\bibitem[\protect\citeauthoryear{{S{\'a}nchez} et~al.,}{{S{\'a}nchez}
  et~al.}{2014}]{Sanchez2014}
{S{\'a}nchez} S.~F.,  et~al., 2014, \mn@doi [\aap]
  {10.1051/0004-6361/201322343}, \href
  {http://adsabs.harvard.edu/abs/2014A%26A...563A..49S} {563, A49}

\bibitem[\protect\citeauthoryear{{Schruba} et~al.,}{{Schruba}
  et~al.}{2011}]{Schruba2011}
{Schruba} A.,  et~al., 2011, \mn@doi [\aj] {10.1088/0004-6256/142/2/37}, \href
  {http://adsabs.harvard.edu/abs/2011AJ....142...37S} {142, 37}

\bibitem[\protect\citeauthoryear{{Schuster}, {Kramer}, {Hitschfeld},
  {Garcia-Burillo}  \& {Mookerjea}}{{Schuster} et~al.}{2007}]{schuster07}
{Schuster} K.~F.,  {Kramer} C.,  {Hitschfeld} M.,  {Garcia-Burillo} S.,
  {Mookerjea} B.,  2007, \mn@doi [\aap] {10.1051/0004-6361:20065579}, \href
  {http://adsabs.harvard.edu/abs/2007A%26A...461..143S} {461, 143}

\bibitem[\protect\citeauthoryear{{Shields}, {Skillman}  \&
  {Kennicutt}}{{Shields} et~al.}{1991}]{Shields1991}
{Shields} G.~A.,  {Skillman} E.~D.,   {Kennicutt} Jr. R.~C.,  1991, \mn@doi
  [\apj] {10.1086/169872}, \href
  {http://adsabs.harvard.edu/abs/1991ApJ...371...82S} {371, 82}

\bibitem[\protect\citeauthoryear{{Spitzer} \& {Baade}}{{Spitzer} \&
  {Baade}}{1951}]{Spitzer1951}
{Spitzer} Jr. L.,  {Baade} W.,  1951, \mn@doi [\apj] {10.1086/145406}, \href
  {http://adsabs.harvard.edu/abs/1951ApJ...113..413S} {113, 413}

\bibitem[\protect\citeauthoryear{{Thilker} et~al.,}{{Thilker}
  et~al.}{2007}]{Thilker2007}
{Thilker} D.~A.,  et~al., 2007, \mn@doi [\apjs] {10.1086/523853}, \href
  {http://adsabs.harvard.edu/abs/2007ApJS..173..538T} {173, 538}

\bibitem[\protect\citeauthoryear{{Thuan}, {Izotov}  \& {Foltz}}{{Thuan}
  et~al.}{1999}]{Thuan99}
{Thuan} T.~X.,  {Izotov} Y.~I.,   {Foltz} C.~B.,  1999, \mn@doi [\apj]
  {10.1086/307877}, \href {http://adsabs.harvard.edu/abs/1999ApJ...525..105T}
  {525, 105}

\bibitem[\protect\citeauthoryear{{Tinsley}}{{Tinsley}}{1980}]{Tinsley80}
{Tinsley} B.~M.,  1980, \fcp, \href
  {http://adsabs.harvard.edu/abs/1980FCPh....5..287T} {5, 287}

\bibitem[\protect\citeauthoryear{{Toribio San Cipriano}, {Garc{\'{\i}}a-Rojas},
  {Esteban}, {Bresolin}  \& {Peimbert}}{{Toribio San Cipriano}
  et~al.}{2016}]{TSC2016}
{Toribio San Cipriano} L.,  {Garc{\'{\i}}a-Rojas} J.,  {Esteban} C.,
  {Bresolin} F.,   {Peimbert} M.,  2016, \mn@doi [\mnras]
  {10.1093/mnras/stw397}, \href
  {http://adsabs.harvard.edu/abs/2016MNRAS.458.1866T} {458, 1866}

\bibitem[\protect\citeauthoryear{{Tremonti} et~al.,}{{Tremonti}
  et~al.}{2004}]{tremonti04}
{Tremonti} C.~A.,  et~al., 2004, \mn@doi [\apj] {10.1086/423264}, \href
  {http://adsabs.harvard.edu/abs/2004ApJ...613..898T} {613, 898}

\bibitem[\protect\citeauthoryear{{Wiegert} \& {English}}{{Wiegert} \&
  {English}}{2014}]{Wiegert14}
{Wiegert} T.,  {English} J.,  2014, \mn@doi [\na]
  {10.1016/j.newast.2013.04.006}, \href
  {http://adsabs.harvard.edu/abs/2014NewA...26...40W} {26, 40}

\bibitem[\protect\citeauthoryear{{Williams}, {Dalcanton}, {Dolphin}, {Holtzman}
   \& {Sarajedini}}{{Williams} et~al.}{2009}]{Williams09}
{Williams} B.~F.,  {Dalcanton} J.~J.,  {Dolphin} A.~E.,  {Holtzman} J.,
  {Sarajedini} A.,  2009, \mn@doi [\apjl] {10.1088/0004-637X/695/1/L15}, \href
  {http://adsabs.harvard.edu/abs/2009ApJ...695L..15W} {695, L15}

\bibitem[\protect\citeauthoryear{{Williams}, {Dalcanton}, {Stilp}, {Dolphin},
  {Skillman}  \& {Radburn-Smith}}{{Williams} et~al.}{2013}]{williams13a}
{Williams} B.~F.,  {Dalcanton} J.~J.,  {Stilp} A.,  {Dolphin} A.,  {Skillman}
  E.~D.,   {Radburn-Smith} D.,  2013, \mn@doi [\apj]
  {10.1088/0004-637X/765/2/120}, \href
  {http://adsabs.harvard.edu/abs/2013ApJ...765..120W} {765, 120}

\bibitem[\protect\citeauthoryear{{Wilman} et~al.,}{{Wilman}
  et~al.}{2005}]{Wilman2005}
{Wilman} D.~J.,  et~al., 2005, \mn@doi [\mnras]
  {10.1111/j.1365-2966.2005.08745.x}, \href
  {http://adsabs.harvard.edu/abs/2005MNRAS.358...88W} {358, 88}

\bibitem[\protect\citeauthoryear{{Wolfe}, {Pisano}, {Lockman}, {McGaugh}  \&
  {Shaya}}{{Wolfe} et~al.}{2013}]{Wolfe2013}
{Wolfe} S.~A.,  {Pisano} D.~J.,  {Lockman} F.~J.,  {McGaugh} S.~S.,   {Shaya}
  E.~J.,  2013, \mn@doi [\nat] {10.1038/nature12082}, \href
  {http://adsabs.harvard.edu/abs/2013Natur.497..224W} {497, 224}

\bibitem[\protect\citeauthoryear{{Yin}, {Hou}, {Prantzos}, {Boissier}, {Chang},
  {Shen}  \& {Zhang}}{{Yin} et~al.}{2009}]{yin09}
{Yin} J.,  {Hou} J.~L.,  {Prantzos} N.,  {Boissier} S.,  {Chang} R.~X.,  {Shen}
  S.~Y.,   {Zhang} B.,  2009, \mn@doi [\aap] {10.1051/0004-6361/200912316},
  \href {http://adsabs.harvard.edu/abs/2009A%26A...505..497Y} {505, 497}

\bibitem[\protect\citeauthoryear{{Zahid}, {Dima}, {Kudritzki}, {Kewley},
  {Geller}, {Hwang}, {Silverman}  \& {Kashino}}{{Zahid} et~al.}{2014}]{zahid14}
{Zahid} H.~J.,  {Dima} G.~I.,  {Kudritzki} R.-P.,  {Kewley} L.~J.,  {Geller}
  M.~J.,  {Hwang} H.~S.,  {Silverman} J.~D.,   {Kashino} D.,  2014, \mn@doi
  [\apj] {10.1088/0004-637X/791/2/130}, \href
  {http://adsabs.harvard.edu/abs/2014ApJ...791..130Z} {791, 130}

\bibitem[\protect\citeauthoryear{{Zaritsky}, {Kennicutt}  \&
  {Huchra}}{{Zaritsky} et~al.}{1994}]{zaritsky94}
{Zaritsky} D.,  {Kennicutt} Jr. R.~C.,   {Huchra} J.~P.,  1994, \mn@doi [\apj]
  {10.1086/173544}, \href {http://adsabs.harvard.edu/abs/1994ApJ...420...87Z}
  {420, 87}

\bibitem[\protect\citeauthoryear{{de Blok}, {Walter}, {Brinks}, {Trachternach},
  {Oh}  \& {Kennicutt}}{{de Blok} et~al.}{2008}]{Blok2008}
{de Blok} W.~J.~G.,  {Walter} F.,  {Brinks} E.,  {Trachternach} C.,  {Oh}
  S.-H.,   {Kennicutt} Jr. R.~C.,  2008, \mn@doi [\aj]
  {10.1088/0004-6256/136/6/2648}, \href
  {http://adsabs.harvard.edu/abs/2008AJ....136.2648D} {136, 2648}

\bibitem[\protect\citeauthoryear{{de Blok} et~al.,}{{de Blok}
  et~al.}{2014}]{Blok2014}
{de Blok} W.~J.~G.,  et~al., 2014, \mn@doi [\aap]
  {10.1051/0004-6361/201423880}, \href
  {http://adsabs.harvard.edu/abs/2014A%26A...569A..68D} {569, A68}

\bibitem[\protect\citeauthoryear{{de Vaucouleurs}, {de Vaucouleurs}, {Corwin},
  {Buta}, {Paturel}  \& {Fouque}}{{de Vaucouleurs}
  et~al.}{1991}]{1991S&T....82Q.621D}
{de Vaucouleurs} G.,  {de Vaucouleurs} A.,  {Corwin} Jr. H.~G.,  {Buta} R.~J.,
  {Paturel} G.,   {Fouque} P.,  1991, \skytel, \href
  {http://adsabs.harvard.edu/abs/1991S%26T....82Q.621D} {82, 621}

\bibitem[\protect\citeauthoryear{{van Zee}, {Salzer}, {Haynes}, {O'Donoghue}
  \& {Balonek}}{{van Zee} et~al.}{1998}]{van_Zee1998}
{van Zee} L.,  {Salzer} J.~J.,  {Haynes} M.~P.,  {O'Donoghue} A.~A.,
  {Balonek} T.~J.,  1998, \mn@doi [\aj] {10.1086/300647}, \href
  {http://adsabs.harvard.edu/abs/1998AJ....116.2805V} {116, 2805}

\makeatother
\end{thebibliography}








\bsp	
\label{lastpage}
\end{document}